\documentclass[twocolumn,prb,aps,floatfix,superscriptaddress,longbibliography]{revtex4-2}

\usepackage{xcolor}
\usepackage{hyperref}

\usepackage{graphicx}
\usepackage{physics}

\usepackage{amsthm}
\usepackage{amsmath}
\usepackage{amssymb}

\newcommand{\angstrom}{\textup{\AA}}

\renewcommand{\vec}[1]{\boldsymbol{#1}}

\usepackage{calligra}
\DeclareMathAlphabet{\mathcalligra}{T1}{calligra}{m}{n}
\DeclareFontShape{T1}{calligra}{m}{n}{<->s*[2.2]callig15}{}

\DeclareSymbolFont{usualmathcal}{OMS}{cmsy}{m}{n}
\DeclareSymbolFontAlphabet{\mathcal}{usualmathcal}

\usepackage{booktabs}

\AtBeginDocument{%
\heavyrulewidth=.08em
\lightrulewidth=.05em
\cmidrulewidth=.03em
\belowrulesep=.65ex
\belowbottomsep=0pt
\aboverulesep=.4ex
\abovetopsep=0pt
\cmidrulesep=\doublerulesep
\cmidrulekern=.5em
\defaultaddspace=.5em
}

\usepackage[color=orange!60,textsize=scriptsize]{todonotes}
\usepackage{siunitx}
\DeclareSIUnit\angstrom{\text{Å}}

\usepackage{soul}

\begin{document}

\title{Localization and Wetting of $^4$He Inside Pre-plated Nanopores}

\author{Sutirtha Paul}
\affiliation{Department of Physics and Astronomy, University of Tennessee, Knoxville, TN 37996, USA}

\author{Taras Lakoba} 
\affiliation{Department of Mathematics \& Statistics, University of  Vermont, Burlington, VT 05405, USA}

\author{Paul E. Sokol}
\affiliation{Department of Physics, Indiana University, Bloomington, IN 47408, USA}

\author{Adrian Del Maestro}
\affiliation{Department of Physics and Astronomy, University of Tennessee, Knoxville, TN 37996, USA}
\affiliation{Min H.~Kao Department of Electrical Engineering and Computer Science, University of Tennessee, Knoxville, TN 37996, USA}

\begin{abstract}
    Low dimensional quantum fluids, where one can probe the effects of enhanced thermal and quantum fluctuations on macroscopic quantum wavefunctions, can be experimentally realized through transverse physical confinement of superfluid helium on scales smaller than the coherence length.  Reaching this scale is difficult, requiring confinement in single or multiple pores with nanometer radii.  Porous silicates such as MCM-41 have a pore radius larger than the coherence length of $^4$He, and in this work we systematically explore the possibility of pre-plating pores with different elements to reduce the pore size without localizing the confined superfluid.  Through a direct solution of the few-body Schrodinger equation combined with quantum Monte Carlo simulations, we explore the behavior of helium confined inside cylindrical nanopores for a range of pre-plating elements, including rare gases and alkali metals. For rare gases, we find that helium remains strongly attracted to the pore walls and any atoms in the core form an incompressible liquid. For alkali metals such as Cs, weak interactions between helium and the pre-plating material prevent localization near the walls and enable delocalization in the pore center.  Our results extend previous results for helium wetting on flat two dimensional coated substrates to the curved geometry inside nanopores, and demonstrate that alkali pre-plated nanopores may enable a tunable one-dimensional confined quantum liquid of helium.  
\end{abstract}

\maketitle

\section{Introduction}

Realisation of Tomonaga-Luttinger liquid behaviour \cite{Tomonaga:1950ak,Luttinger:1963gd,Mattis:1965iq, Haldane:1981eh, Giamarchi:2004bk} in physical systems requires transverse spatial confinement on a scale smaller than the coherence length of their quantum many-body wavefunction.  This has been achieved in spin chains \cite{Lake:2005ho}, quantum nanowires \cite{Jompol:2009sc}, ultracold atoms \cite{Senaratne:2022fg} and confined superluids \cite{DelMaestro22} with signature features of quantum hydrodynamics in one dimension (1D), such as spin charge separation, being experimentally reported. A persistent goal has been the engineering of high density 1D quantum liquids with tunable strong interaction effects. A promising platform in this regard is the quantum liquid helium-4 confined inside single engineered nanopores 
\cite{Savard:2009fc, Savard:2011hd, Velasco:2012de, Velasco:2014fe, Duc:2015sa, Botimer:2016pd} or in ordered porous media \cite{Wada:2001hh,Yamamoto:2004ss,Taniguchi:2005pr, Toda:2007sf, Ikegami2005, Azuah2013,Prisk:2013br,Yager:2013cva,Demura:2015hq,Demura:2017zj,Bryan:2017gz,Taniguchi_2018,Bossy:2019,Taniguchi2020,DelMaestro22,Kuribara:2023sf}.  The bulk superfluid transition temperature of $^4$He is $T_\lambda \simeq \SI{2.17}{\kelvin}$ and below this temperature, the superfluid coherence length drops rapidly to a length scale $\xi \sim \SI{1}{\nano\meter}$ fixed by the microscopic details of a vortex core \cite{Galli:2014uw, Amelio:2018cx}.  Numerical simulations of confined helium inside cylindrical pores \cite{Chakravarty:1997eq, Cole:2000tz, Gordillo:2000qo, Rossi:2006cg, DelMaestro:2011et,DelMaestro:2012td,Kulchytskyy:2013qq,Markic:2015bu,Markic:2018bw,Markic:2020,Nava:2022hk} have identified a microscopic structure of nested cylindrical shells driven by the interplay between helium-helium and helium-substrate interactions.  When the radii of the confining media enters the nanoscale regime, a central 1D core of atoms can arise with low energy behavior described by the Luttinger liquid (LL) framework with a LL parameter $K$ that is tunable through the density of the quantum liquid.  Reaching this scale through physical confinement in porous silicates such as MCM-41 \cite{Kresge:1992vv, Sonwane:1999xl} and FSM-16 \cite{Inagaki:1996mc,Toda:2007sf} has been challenging due to their ``as synthesized'' radii of $\gtrsim \SI{2}{\nano\meter}$. This is still too large to directly probe the 1D limit and only quantitative deviations from 3D bulk superfluid behavior have been observed \cite{Prisk:2013br,Bryan:2017gz}.  

Recently, it was theoretically proposed \cite{Nichols:2020of} that a reduction in pore radius could be achieved through pre-plating of porous media with an adsorbed noble gas.  This was subsequently explored experimentally by adsorbing a single layer of Ar into MCM-41 \cite{DelMaestro22}, reducing the effective pore radius to $\sim \SI{1.5}{\nano\meter}$.  Inelastic neutron scattering measurements of helium confined inside Ar pre-plated MCM-41 demonstrated signatures of LL behavior; however, the density of the 1D liquid of $^4$He formed at the center of the nanopores was found to be mostly insensitive to the external vapor pressure of helium gas surrounding the sample.  This behavior has been confirmed by subsequent measurements in the same system and is hypothesized to be a result of disorder in the pores combined with the strong van der Waals interactions between He and the pre-plating material Ar \cite{Nichols:2020of}.  

\begin{figure}[t]
    \centering
    \includegraphics[width=\columnwidth]{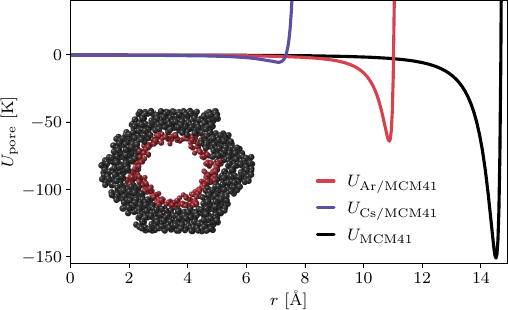}    
    \caption{The radial confinement potential (in units of $k_{\rm B}$) of a helium atom inside a bare MCM-41 nanopore, and those pre-plated with Ar and Cs. The inset shows a snapshot of Ar atoms (red) adsorbed inside a single pore (grey) computed with molecular dynamics.  The potentials were computed via the method reported in Ref.~\cite{Nichols:2020of} and demonstrate the ability of tuning both the magnitude and effective scale of confinement via pre-plating.
    \label{fig:PotentialcompMCM}}
\end{figure}

It is thus natural to consider the effects of different pre-plating materials on helium adsorbed inside ordered porous materials. In this paper, we move beyond the noble gases and theoretically consider a range of pre-plating elements including alkali, alkaline earth, and transition metals.  Our intuition is guided by a large body of research on helium wetting of two dimensional solid surfaces \cite{Graham:1989qh, Rutledge:1992, Cheng:1993, Cheng:1993ri, Treiner1993,Vidali:1991,Taborek1994,VanCleve2007,Gatica:2009yo} including the well-known experimental result that helium fails to wet the surface of cesium, instead forming a phase-separated state of discrete $^4$He droplets at the interface \cite{Nacher:1991,Rutledge:1992,Ketola:1992,Mukherjee:1992}. The ability to manipulate both the radii and strength of the confinement potential for different pre-plating materials  can be seen in Fig~\ref{fig:PotentialcompMCM}. 

We proceed by first solving the Schr{\"o}dinger equation via an efficient relaxation method \cite{Lakoba:2024} for two interacting helium atoms confined within idealized cylindrical cavities of varying radii carved inside an infinite medium composed of different materials.  Our results show clear differences in $^4$He localization behavior as the material of the surrounding medium is modified, which we quantify through an analysis of the two-particle wavefunction inside the pore. We find that for the alkali metals Cs, K, and Rb, the density of helium atoms in the center of the core remains appreciable while for Au and the rare gases Ne and Ar, it is vanishingly small. In these cases, the He atoms remain localized near the walls of the cylindrical pore.  Mg represents an intermediate case, exhibiting both types of behavior.  As the number of $^4$He atoms inside the pore is increased beyond two, we study the system with finite temperature grand canonical path integral quantum Monte Carlo \cite{Boninsegni:2006ed, Boninsegni:2006gc, pimcrepo} exploring the density and compressibility of an emergent 1D core of atoms as a function of external pressure below $T_\lambda$. 

The main finding of this study is that in the many-particle limit relevant for experiments, alkali metal pre-plated pores support a compressible quasi-1D core for a range of core radii opening up the possibility of pressure tuning the interaction strength in confined quantum liquids of helium. The results for all pre-plating materials considered are summarized in a phase diagram (see Fig.~\ref{fig:Phase}) 
that extends previous results for $^4$He wetting on 2D flat surfaces \cite{Treiner1993} to atoms confined inside a cylindrical geometry.

The remainder of the paper is organized as follows. We begin with a definition of the microscopic confinement model for helium inside pre-plated nanopores and then briefly describe the numerical relaxation method we use to exactly solve the two-particle Schr{\"o}dinger equation.  An analysis of the resulting ground state densities of confined helium yields our proposed phase diagram as a function of pore radius for different plating materials.  We then study the many-body limit and report on the structure and properties of helium inside the pore.  We conclude with a discussion on the near-term experimental implications of our findings.  Appendices include information on the accuracy and performance of our numerical methods.

\section{Helium-4 inside Pre-Plated Cylindrical Nanopores}
\label{sec:2}
\begin{figure}
    \centering 
    \includegraphics[width=\columnwidth]{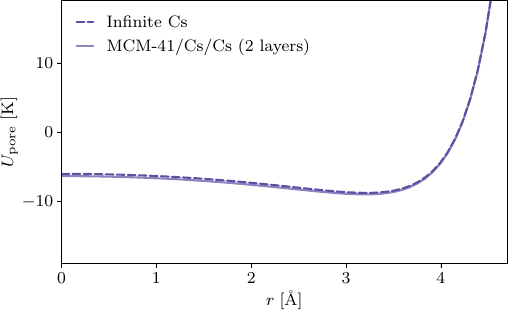}    
    \caption{Accuracy of the effective potential in Eq.~\eqref{eq:Vpore} (in units
        of $k_{\rm B}$) corresponding to a cylindrical cavity in an infinite
        medium of Cs compared to a superposition of MCM-41 and two layers of
        Cs. \label{fig:U}}
\end{figure}

We consider a model system of $N$ $^4$He atoms of mass $m$ and positions $\vec{r}_i = (x_i,y_i,z_i)$ interacting through a potential $V$ \cite{Przybytek:2010js,Cencek:2012iz} confined inside a single smooth cylindrical pore $U_{\rm pore}$ with periodic boundary conditions along the long $z$-axis 
described by the Hamiltonian
\begin{equation}
    H = -\frac{\hbar^2}{2m}\sum_{i=1}^N \nabla_i^2 + \sum_{i=1}^N U_{\rm
    pore}(\vec{r}_i, R) + \frac{1}{2}\sum_{i,j} V(\vec{r_i}-\vec{r_j}).
\label{eq:Ham}
\end{equation}
\begin{table}[t]
    \renewcommand{\arraystretch}{1.5}
    \setlength\tabcolsep{5pt}
    \begin{tabular}{@{}lllll@{}} 
	\toprule
	Preplating element& $\sigma (\angstrom)$ & $\varepsilon$ (K) & n ($\angstrom^{-3}$)& Reference \\
	\midrule
	Argon (Ar)  & 3.0225 & 36.136 & 0.0265 & \cite{DelMaestro22,Arblaster2018} \\
	Cesium (Cs) & 5.44  & 1.359 & 0.0091 & \cite{Treiner1993,Arblaster2018} \\
	Gold (Au) &  3.305 & 19.59 &  0.0595 & \cite{Vidali:1991,Arblaster2018} \\
	Magnesium (Mg)   &3.885& 5.661&  0.0437& \cite{Treiner1993,Arblaster2018} \\
	Neon (Ne) &  2.695 	& 19.75 & 0.0440 & \cite{DelMaestro22,Muser1995,Batchelder1967}\\
	Potassium (K) & 5.14  & 1.512 &0.0139 & \cite{Treiner1993,Arblaster2018}\\
	Rubidium (Rb) & 5.417 & 1.251 &0.0114& \cite{Treiner1993,Arblaster2018}\\
	\bottomrule
\end{tabular}
\caption{The simulation parameters used. The first or first two references are for the L-J parameters while the last is for the density. The parameters given here are for those of He-material interaction calculated by Berthelot mixing from the individual parameters \cite{Boda:2008}.}
\label{tab:eps-sig}
\end{table}
For $^4$He inside pre-plated MCM-41, the potential $U_{\rm pore}$ was computed
in Ref.~\cite{Nichols:2020of} incorporating a single layer of pre-plating
material inside a cylindrical cavity of radius $R$ and length $L$.  In this
work, we exploit the fact that porous materials synthesized via a surfactant
templating admit variable radii with length:radius aspect ratios of $\sim
1000:1$ and can admit multiple layers of adsorbent.  The layers act to screen the silicate potential and we can instead consider an idealized system of a cavity inside an infinite medium of the pre-plating material where $U_{\rm pore}$ is taken to be of Lennard-Jones type \cite{Tjatjopoulos:1988jl,Zhang:2004xo}: 
\begin{equation}
    U_{\rm pore} (r,R) =
    \frac{\pi n \varepsilon\sigma^3}{3}\qty[\qty(\frac{\sigma}{R})^9\!\!
    \mathop{u_9}\qty(\tfrac{r}{R}) - \qty(\frac{\sigma}{R})^3\!\!
    \mathop{u_3}\qty(\tfrac{r}{R})]
    \label{eq:Vpore}
\end{equation}
with
\begin{align*}
u_9(x) &= \frac{1}{240(1-x^2)^9} \bigl[ \\
    &(1091 + 11156x^2 + 16434x^4 + 4052x^6 + 35x^8)E(x) \\
    &- 8(1-x^2)(1+7x^2)(97+134x^2+25x^4)K(x)\bigr]  \\
u_3(x) &= \frac{2}{(1-x^2)^3} \qty[(7+x^2)E(x) - 4(1-x^2)K(x)]
\end{align*}
where $r$ is the distance of an atom from the axis of the cylinder. The density of the media $n$, the strength of the interaction $\varepsilon$ and the hard-core distance $\sigma$ are detailed in Table~\ref{tab:eps-sig} for different pre-plating materials. $K(x)$ and $E(x)$ are the complete elliptic integrals of the first and second kind.  The accuracy of this simplified model can be seen in Fig.~\ref{fig:U}
where only small quantitative deviations are observed for multiple layers of the weakest pre-plating materials. 

\section{Simulation Details}

Our approach to model a system governed by Eq. \eqref{eq:Ham}
is to first solve the Schr\"odinger equation for $N=2$ helium
atoms subject to confinement by different pre-plating materials before extending to the many-body limit via quantum Monte Carlo simulations.

\subsection{$N=2$: Relaxation Method}

For $N = 2$ particles, we can exactly determine the ground state wavefunction $\psi\qty(\vb*{r}_1,\vb*{r}_2)$ of Eq.~\eqref{eq:Ham} via an accelerated relaxation method, which iteratively solves the Schr\"odinger equation beginning from a trial wavefunction. To improve the convergence rate, we use a pre-conditioner $\hat{P}^{-1}$ to damp higher Fourier modes, combined with explicit slow-mode elimination at each iteration step \cite{Lakoba:2007} resulting in fast relaxation to the ground state.  For the particular system under consideration, we found that even a non-accelerated relaxation method is faster than a direct implementation of the Lanczos method.  Defining $\psi_n$ as the normalized value of the wavefunction at iteration step $n$, the next value is given by the following algorithm:
\begin{equation}
    \tilde{\psi}_{n+1} =  \psi_n
    - \delta t \, I_n
    \label{eq:item-0a}
\end{equation}
where:
\begin{equation}
    I_n = \hat{P}^{-1}\hat{L}_0 \psi_n  
    - \qty(\!1 - \frac{s}{\delta t}   \frac{\braket{\phi_n}{\hat{P}\phi_n}}{\braket{\phi_{n}}{\hat{L}_0\phi_{n}}})
    \frac{\braket{\phi_n}{\hat{L}_0\psi_n}}{\braket{\phi_n}{\hat{P}\phi_n}}\phi_n\,,
\label{eq:In}
\end{equation}
\begin{equation}
    \hat{L}_0 \psi_n \equiv \bar{H} \psi_n - \frac{\braket{\bar{H} \psi_n}{\hat{P}^{-1}\psi_n}}{\braket{\psi_n}{\hat{P}^{-1}\psi_n}}\psi_n\,,
\label{eq:L0}
\end{equation}
\begin{equation}
    \phi_n = \psi_n - \psi_{n-1}\,;
\label{eq:phi}
\end{equation}
and then
\begin{equation}
    \psi_{n+1} = \tilde{\psi}_{n+1} / \sqrt{\braket{\tilde{\psi}_{n+1}}{\tilde{\psi}_{n+1}}}\,.
\label{eq:psinplus1}
\end{equation}
Above, the pre-conditioner is
\begin{equation}
    \hat{P}^{-1} = \frac{1}{\bar{E}_{\rm cut} - r_m^2\nabla^2} 
    \label{eq:Pinv}
\end{equation}
with $\bar{E}_{\rm cut} = E_{\rm cut}/E_{0}$, where $E_{\rm cut}$ is an energy cutoff \cite{Lakoba:2024} chosen to optimize the acceleration provided by $\hat{P}$. We work with dimensionless quantities by picking the $^4$He interaction minimum, $r_{m} = \SI{2.9673}{\angstrom}$ as our characteristic length scale and $E_{0} = \hbar^2/(2m r_{m}^2) \simeq \SI{0.7}{\kelvin}$ as our characteristic energy scale (i.e.\@ $\bar{H} = H/E_0$).
In Eq.~\eqref{eq:L0}, the fraction on the r.h.s. is the estimate of the energy, $\bar{E}_n$, at the $n^{\rm th}$ iteration; so for the exact wavefunction $\psi$ one would have $\hat{L}_0\psi=0$.  
When implementing $\bar{H}$ in Eq.~(\ref{eq:L0}), we cut off both the confinement $U$ and interaction $V$ potentials in Eq.~(\ref{eq:Ham}): $\max_{\vec{r}_1,\vec{r}_2} U,V = E_{\rm cut}$, with $E_{\rm cut}$ being sufficiently high to not significantly change the physics of the system and yet significantly low to not impede convergence of the iterations; we found $E_{\rm cut}=\SI{300}{\kelvin}$ to be suitable. In Eq.~\eqref{eq:Pinv}, the operator is implemented in momentum space using the Fast Fourier Transform (see Appendix~\ref{app:ConDetails}).  The parameters $\delta t$ and $s$ are chosen empirically to ensure an optimum convergence rate. For potentials with shallow well depths ($\varepsilon/E_0 \le 15$)  we use $\delta t = 1.2$ while for deeper potentials ($\varepsilon/E_0 > 15$) a smaller $\delta t = 0.8$ is required. As in \cite{Lakoba:2024}, we also found that $s=0.1$ yields the fastest convergence (attributed to the interaction having a hard-core). Further details on convergence and accuracy are given in Appendix~\ref{app:ConDetails}. The results obtained by this method for all pre-plating materials studied are provided in the next section. 

\begin{figure*}[t]
\begin{center}
   \includegraphics[width=0.99\linewidth]{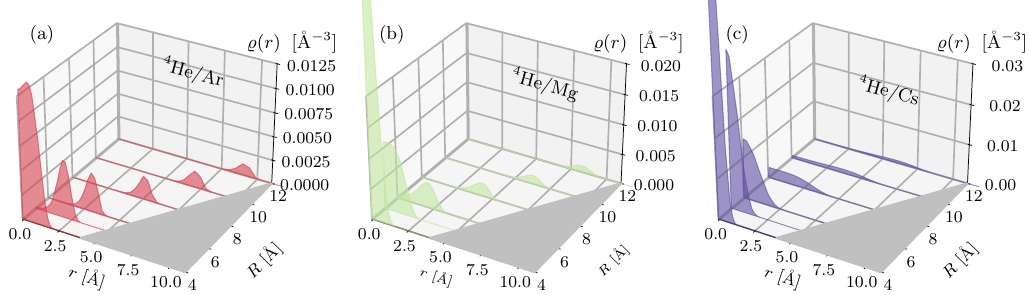}
\end{center}
\caption{The two body radial density $\rho(r)$ of $N=2$ $^4$He atoms confined inside nanopores of different radii for three different pre-plating elements: (a) Ar, (b) Mg, and (c) Cs. The shaded triangular region on the base of the plots indicates excluded volume due to the pore wall; the pore center is at $r=0$. The data as $r\to 0$ has been clipped for comparison due to the radial normalization condition on the density (Eq.~\eqref{eq:norm4rho}).}
\label{Fig:Waterfalls}
\end{figure*}

\subsection{$N>2$: Path Integral Quantum Monte Carlo}
\label{subsec:pimc}
The relaxation method above allows us to determine the exact ground state $\psi\qty(\vb*{r}_1,\dots,\vb*{r}_N)$ on a $(M^{3N})$-dimensional spatial grid (where $M$ is the number of grid points), but becomes computationally prohibitive for $N>2$. In this many-body regime, we instead use path integral quantum Monte Carlo \cite{Ceperley:1995gr, Boninsegni:2006ed, Sarsa:2000jl,pimcrepo} which provides stochastically exact results for ground state expectation values $\expval{\mathcal{O}}_0$ as well as finite temperature observables in the grand canonical ensemble 
\begin{equation}
    \expval{{\mathcal{O}}}  = \frac{1}{\mathcal{Z}} \Tr[\mathcal{O}\ \mathrm{e}^{-\beta {(H-\mu N)}}]\, .
\label{eq:thermal_expectation_value}
\end{equation}
Here $\beta = 1/T$ (we work in units where $k_{\rm B} = 1$), $\mu$ is the chemical potential, $\mathcal{Z} = \Tr \mathrm{e}^{-\beta {(H-\mu N)}}$, and $N$ is the particle number operator.  This method has been extensively used to study the behavior of $^4$He quantum liquids under confinement at low temperatures \cite{DelMaestro:2011et,Kulchytskyy:2013qq,Nichols:2020of, Nava:2022hk} and can provide efficient access to detailed structural and emergent properties for the geometry under consideration.

\section{Simulation results}

\subsection{2-Particle Densities in Pre-Plated Nanopores}

\begin{figure}[h]
\begin{center}
   \includegraphics[width=\columnwidth]{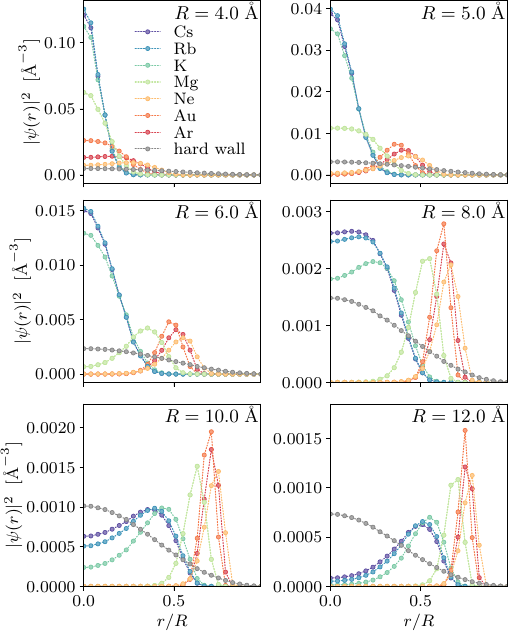}
\end{center}
\caption{The radial density distribution of two helium atoms confined inside nanopores with different pre-plating materials and different radii.  ``Hard wall'' indicates a confinement potential with $U(r<R) = 0$ and $U(r\ge R)=E_{\rm cut}$.} 
\label{fig:rhoDensity}
\end{figure}

For $N=2$, we determined the ground state wavefunction for $^4$He atoms confined inside cylindrical nanopores of varying radii from $R = \qtyrange{4}{12}{\angstrom}$ for all pre-plating materials listed in Table~\ref{tab:eps-sig}. Utilizing translational symmetry in the $z$-direction, we define a relative coordinate $z = z_2-z_1$ and compute the single particle number density in Cartesian coordinates as: 
\begin{equation}
\rho(x,y,z) = \qty|\psi(x,y,z)|^2 = \!\int\!\! \dd{x_2}\! \int\!\! \dd{y_2}|\psi(x_,y,x_2,y_2,z)|^2\, .
\end{equation}
It is more natural to work in cylindrical coordinates $(r,\theta,z)$ where $r$ is the defined to be the distance from the axis of the cylinder. This needs to be done with some care as a direct coordinate conversion can become numerically unstable.  Instead, we rebin the Cartesian density in radial coordinates by setting a radius of influence, before averaging over the azimuthal angle $\theta$ and $z$ to obtain: 
\begin{equation}
    \varrho(r) = \qty|\psi(r)|^2 = \frac{1}{2\pi L} \int_0^L \dd{z} \int_0^{2\pi} \dd{\theta} |\psi(r,\theta,z)|^2
\end{equation}
such that 
\begin{equation}
    2\pi L\int_0^R r \dd{r} \varrho(r) = 2 \, (=N).
    \label{eq:norm4rho}
\end{equation}
Figure \ref{Fig:Waterfalls} shows the radial density of two helium atoms confined inside pores pre-plated with Ar, Mg, and Cs, for different radii.  
For argon, $^4$He atoms are strongly bound to the pore wall and as the radius of the pore is increased, they are pulled away from the pore center.  Cesium represents the opposite behavior, with atoms instead delocalizing near the pore axis in order to minimize their kinetic energy while maintaining $\rho(r\to0) \ne 0$. Magnesium pre-plating appears to be more Cs-like for small radii, but eventually behaves similar to Ar as $R \ge \SI{8}{\angstrom}$. The full data set (for all radii and pre-plating materials) is shown in Fig.~\ref{fig:rhoDensity} where we have included a ``hard wall'' for comparison where the confinement potential is set to be zero for $r<R$ and to $E_{\rm cut}$ for $r \ge R$.
%

\subsection{Wetting inside Nanopores}
\label{subsec:wetting}

\begin{figure}[h!]
\begin{center}
   \includegraphics[width=\columnwidth]{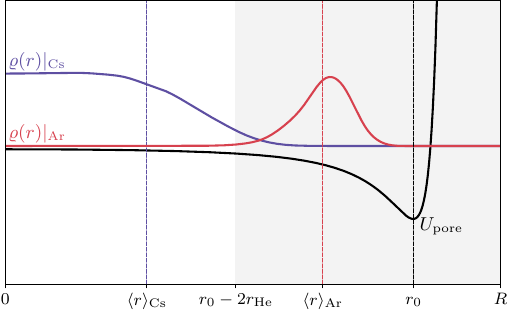}
\end{center}
\caption{The radial density of $^4$He inside a nanopore with $R=\SI{8}{\angstrom}$ (scale is arbitrary) for two different pre-plating materials Ar and Cs demonstrating the effect of the two wetting parameters defined in Eqs.~\eqref{eq:Oaver} and \eqref{eq:Ofwell}.} 
\label{fig:wettingOP}
\end{figure}

To characterize and quantify the adsorption/wetting behaviour observed for different pre-plating materials, we identify two different wetting order parameters:  
\begin{align}
    \label{eq:Oaver}
    \expval{\frac{r}{R}} &= \frac{2\pi L}{N R}\int_0^R r^2 \dd{r} \varrho(r) \\
    \label{eq:Ofwell}
    f_{\text{well}} &= \frac{2\pi L}{N} \int_{r_0 - 2r_{\rm He}}^{R} r\dd{r} \varrho(r)\, ,
\end{align}
where $r_0$ is the location of the minimum of $U_{\rm pore}$, and $r_{\rm He} = \SI{1.4}{\angstrom} \approx r_m/2$ is the van der Waals radius of He.

\begin{figure}[t]
    \includegraphics[width=\columnwidth]{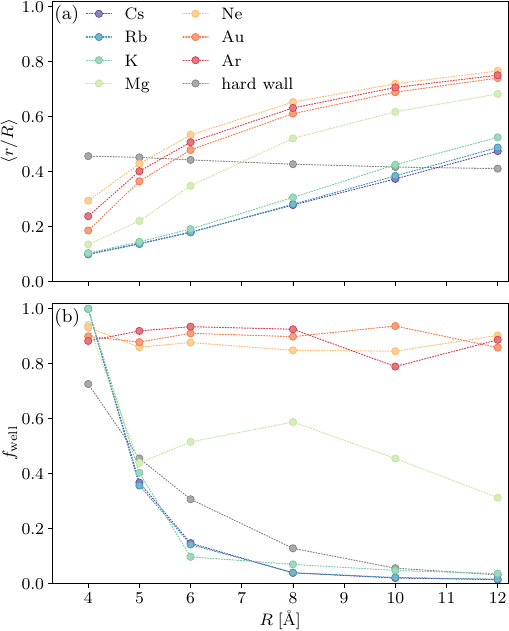}    
    \caption{The evolution of two different wetting order parameters as a function of the pore radius $R$ for the various pre-plating materials considered.  Panel (a) shows the average radial position of the $N=2$ helium atoms inside the pore, while panel (b) shows the fraction $f_{\rm well}$ of particles localized in the attractive potential well near the pore wall. For both wetting parameters, clear distinction between alkali metals and rare gases can be seen, with the former displaying non-wetting behavior. $r_0 = R$ for the hard wall.} 
\label{fig:averfwell}
\end{figure}

As can be seen schematically in Fig.~\ref{fig:wettingOP}, Eq.~\eqref{eq:Oaver} captures the average radial position of $^4$He atoms inside the nanopore. For non-wetting materials $\expval{r}$ should be closer to the center of the pore ($r=0$) than to the wall ($r=R$) and thus we would expect $\expval{r/R}$ to be small for weak adsorbers like Cs and Rb and closer to unity for strong adsorbers like Ar and Ne.  This works for larger radii nanopores, but fails to strongly discriminate between pre-plating materials for small $R$ where all $^4$He atoms are close to the center as can be seen in Fig.~\ref{fig:averfwell}(a). 
The fraction of particles $f_{\rm well}$ in the potential well $U_{\rm pore}$ is shown in Fig.~\ref{fig:averfwell}(b) where we again see clear separation with non-wetting behavior (characterized by vanishing density in the well) for $R\ge\SI{8}{\angstrom}$ for the alkali metals Cs, Rb, and K, and strong well localization and wetting for Ne, Au and Ar. Again, Mg displays intermediate behavior.

\begin{figure}[h]
\begin{center}
   \includegraphics[width=\columnwidth]{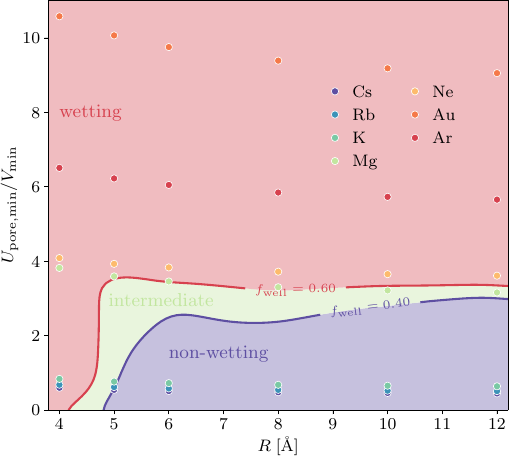}
\end{center}
\caption{A schematic wetting phase diagram for $^4$He inside pre-plated nanopores as a function of adsorption potential well depth (in units of the well depth of the helium-helium interaction) and the pore radius. The decision boundaries for non-wetting and wetting are shown as contour labels.  The phase boundaries were determined by linear interpolation of the sparse data.}
\label{fig:Phase}
\end{figure}

In analogy to the wetting phase diagram for helium on flat 2D surfaces in Ref.~\cite{Treiner1993} we can construct a similar phase diagram for quantum wetting using $f_{\text{well}}$ as the discriminator with the results shown in Fig.~\ref{fig:Phase}.
This figure represents a major result of our paper. We find that (for $N=2$ $^4$He atoms), there is a minimal radius ($R \approx \SI{5}{\angstrom}$) and minimal pre-plating adsorption strength ($U_{\rm pore, min}/V_{\rm min} \approx 2$, with the helium-helium interaction energy minimum being $V_{\rm min} \simeq \SI{10}{\kelvin}$) where a finite density of helium atoms remains in the nanopore center, delocalized from the pore wall.  Based on our results, using an alkali metal (e.g. Cs) as the pre-plating material is the most promising candidate for a tunable low-dimensional quantum liquid of $^4$He.  

\subsection{Quantum Monte Carlo: Filling the Nanopores}
\label{subsec:qmcfilling}

To examine the validity of this prediction in the many-body regime and at finite temperature we have performed extensive quantum Monte Carlo simulations focusing on two pre-plating materials which are representative of the dominant wetting behaviors observed in Fig.~\ref{fig:Phase}: Ar and Cs.

\begin{figure}
    \includegraphics[width=\columnwidth]{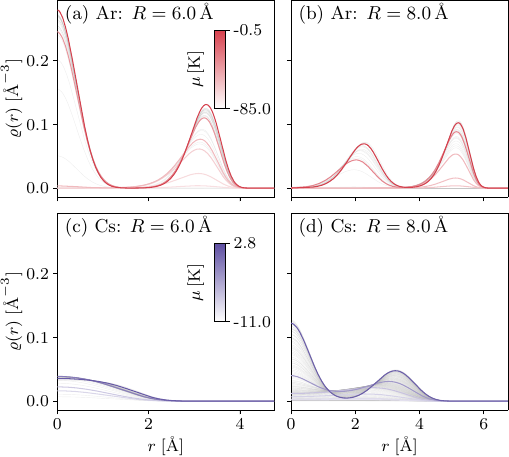} 
    \caption{The radial density as a function of position inside the nanopore computed via quantum Monte Carlo for different chemical potentials $\mu$ at $T = 2 K$ and $L=\SI{25}{\angstrom}$. The panels correspond to different elements and pore radii as indicated.  The minimum and maximum number of particles for each case $(\expval{N}_{\rm min},\expval{N}_{\rm max})$ are: (a) (0,60), (b) (0,106), (c) (1,8), and (d) (1,43) where we have rounded all values to the nearest integer. Quantum Monte Carlo error bars very small due to the self-averaging and are not included in this figure.} 
    \label{fig:PIMC-densities}
\end{figure}

We focus on two estimators: the average number of particles inside the pore $\expval{N}$ and the radial density:
\begin{equation*}
\varrho(r) = \expval{\sum_i^N \delta\qty(\sqrt{x_i^2+y_i^2}-r)}\, .
\end{equation*}
Figure~\ref{fig:PIMC-densities} shows the radial density as we fill up the nanopores by increasing the chemical potential $\mu$ for $R=\SIlist{6,8}{\angstrom}$ below the superfluid transition at $T=\SI{2.0}{\kelvin}$ for $L=\SI{25}{\angstrom}$.  Following Ref.~\cite{Nichols:2020of} we choose an imaginary time step of $\tau = \SI{0.004}{\kelvin^{-1}}$ to ensure that any systematic imaginary time discretization error is smaller than the stochastic uncertainty in our results.  For Ar, shown in panels (a-b), the average number of particles is zero for the lowest chemical potentials considered, and is over 100 for $R=\SI{8}{\angstrom}$ and $\mu = \SI{-0.5}{\kelvin}$.
The behavior of $^4$He inside the pore is analogous to the $N=2$ case, with atoms first becoming localized near the pore wall.  As the chemical potential is increased and additional atoms enter the pore, helium-helium interactions stabilize a transition to a state with a single core of atoms at the axis of the cylinder surrounded by a cylindrical shell (as in panel (a)) or two shells (as in panel (b)) depending on commensuration effects between $r_{\rm He}$ and $R$ (i.e. interactions set the radial separation between the integer number of core plus shells).  This behavior is fully consistent with previous simulation studies for strongly attractive pore walls \cite{Nichols:2020of, Nava:2022hk}. 

For the Cs plated nanopores in panels (c-d), we see very different behavior. For both $R=\SI{6}{\angstrom}$ and $R=\SI{8}{\angstrom}$, a finite density of $^4$He atoms remains in the center of the cylinder. For $R=\SI{8}{\angstrom}$ the central core is surrounded by a shell which remains separated from the pore wall. This configuration allows for a smooth change in the number of particles in the central core as indicated by the density of curves for different values of $\mu$.  Thus, our many-body findings are consistent with the phase diagram presented in Fig.~\ref{fig:Phase}.

\begin{figure}
\centering
\includegraphics[width=\columnwidth]{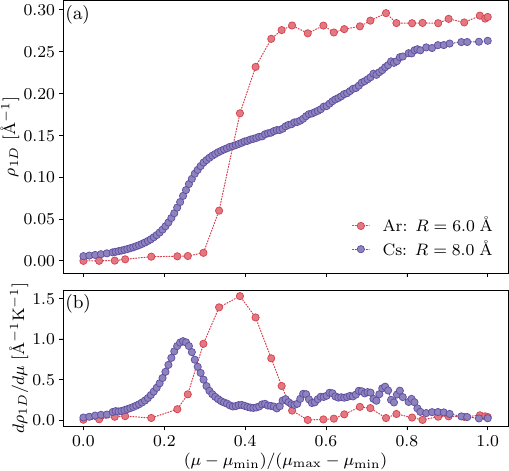}
\caption{Details of the 1D core in pre-plated nanopores. (a) $\rho_{1D}$ as a function of rescaled chemical potential where $\mu_{\rm min/max}$ are given in Fig.~\ref{fig:PIMC-densities}. Errorbars are smaller than the symbol size.  (b) The 1D compressibility obtained as a discrete derivative of the density data in (a) utilizing a Guassian filter with $\sigma=2$ to smooth the data. D}
\label{fig:rho1dvsmu}
\end{figure}

To quantitatively confirm that for Cs coated nanpores the central core of atoms remains compressible and thus tunable by external pressure, we define the 1D core density (see Ref.~\cite{Nichols:2020of} for more details) as:
\begin{equation}
    \rho_{1D} = 2\pi \int_0^{R^{(1)}_{\rm min}} r  \dd{r} \varrho(r)
\end{equation}
where $R^{(1)}_{\rm min}$ is the radius of the first minimum in $\varrho(r)$ for the completely filled pore (corresponding to $\mu_{\rm max}$). Using the data in Fig.~\ref{fig:PIMC-densities} we extract $R^{(1)}_{\rm min} \simeq \SI{1.68}{\angstrom}$ for Ar and $R=\SI{6.0}{\angstrom}$ and $R^{(1)}_{\rm min} \simeq \SI{1.72}{\angstrom}$ for Cs and $R=\SI{8.0}{\angstrom}$. These combinations of element and radius were chosen as they include both a core and surrounding shell.  The resulting 1D core densities are shown in Fig.~\ref{fig:rho1dvsmu}(a).
Here we have rescaled the $x$-axis to enable direct comparison of the two pre-plating materials.  As previously reported for Ar pre-plating \cite{Nichols:2020of}, we observe a single well-defined step from the empty to fully filled core, which remains at constant density over a wide range of chemical potentials.  For Cs pre-plating, the behavior is qualitatively different, with a smoother step in density followed by an extended region of continuous growth in filing which finally saturates at a density of $\rho_{1D}\simeq \SI{0.25}{\angstrom^{-1}}$ for $\mu_{\rm max} = \SI{2.8}{\kelvin}$. This is slightly below the saturation density for Ar pre-plating which occurs at $\mu_{\rm max} = \SI{-0.5}{\kelvin}$.  This maximal 1D density has been observed in previous simulations of confined $^4$He \cite{DelMaestro:2012td,Nichols:2020of,Rosenow:2024fo} and is mostly independent of the details of $U_{\rm pore}$, arising instead from the interaction potential between $^4$He atoms in the 1D core. It can be sensitive to $L$ on the order of \SI{0.02}{\angstrom^{-1}} for short pores due to finite size effects in the central core \cite{Nichols:2020of}.  A numerical derivative of $\rho_{\rm 1D}$ yields the compressibility of the 1D core as shown in Fig.~\ref{fig:rho1dvsmu}(b).  While both pre-plating materials are compressible during initial filling, signified by broad peaks, only the Cs coated pore remains compressible over a range of $\mu$.

\section{Discussion}

Our results provide a systematic framework for understanding how pre-plating with different elements modifies the microscopic structure of MCM-41 porous silicate and may lead to emergent quantum phases of helium confined within cylindrical nanopores. By combining an exact two-body solution with many-body quantum Monte Carlo simulations, we have demonstrated that the interplay between helium–substrate interactions and geometric confinement leads to sharply contrasting wetting regimes. Rare gases and noble metals such as Ar, Ne, and Au act as strong adsorbers, attracting helium to the pore walls and suppressing the formation of a compressible one-dimensional (1D) core. In contrast, alkali metals (Cs, Rb, K) generate weakly attractive or even non-wetting coatings, stabilizing a finite central density of helium and opening the possibility of a tunable Tomonaga–Luttinger liquid via external pressure.

The emergence of a compressible 1D helium core in Cs-plated nanopores is potentially significant. While Ar pre-plating had previously been shown to reduce the effective pore radius and reveal features of Luttinger liquid physics \cite{Nichols:2020of, DelMaestro22}, the resulting core density remained largely insensitive to external pressure. Our findings suggest that replacing Ar with Cs addresses this situation: the non-wetting boundary condition allows for a continuous tuning of the 1D density with chemical potential, enabling control over the Luttinger parameter $K$. This finding underscores the role of surface chemistry in realizing low-dimensional quantum fluids through physical confinement and offers a direct pathway to engineer interaction strength via pressure, rather than relying solely on geometry.

The wetting phase diagram we construct extends prior flat two-dimensional surface studies into the cylindrical geometry relevant for nanopores. The existence of a threshold in both pore radius and adsorption strength highlights the competition between minimizing either the kinetic energy through delocalization in the pore center, or the potential energy by freezing near the wall.  This balance explains the intermediate behavior observed for Mg, which exhibits a crossover from Cs-like to Ar-like confinement depending on pore radius. The framework developed here provides a predictive tool for selecting candidate pre-plating materials for experiments seeking realizations of 1D quantum liquids.

Experimentally, our predictions are immediately testable. High-resolution inelastic neutron scattering of Cs-coated MCM-41 nanopores could probe the density profile and dynamic structure factor of confined helium, directly testing the predicted compressible 1D core through the excitation spectrum.  Moreover, transport measurements—such as pressure-driven flow or torsional oscillator studies—could reveal nontrivial scaling of the superfluid response with density, offering a direct window into Luttinger liquid behavior. The pressure-tunable density we identify may also allow experimentalists to explore the crossover between weak and strong interaction regimes in low dimensional superfluids.  While we have included results for the 1D density as a function of chemical potential (the external control \emph{knob} in our quantum Monte Carlo simulations), we can estimate the experimentally accessible pressure regime by exploiting the virial equation of state up to second order using the known temperature dependence of the second coefficient $B_2(T)$ for bulk $^4$He at saturated vapor pressure \cite{Donnelly:1998}.  The conversion yields a relevant pressure range between 0.01 and 30~atm.

Looking forward, several open questions remain. The role of disorder, present in real porous silicates, is likely to interplay with the non-wetting conditions enabled by coating with alkali metals. Whether Cs pre-plating can smooth wall roughness in longer pores to suppress localization is an important avenue for future theoretical and experimental studies.  Furthermore, our phase diagram was computed for only $N=2$ atoms and while the many-body results we present are fully consistent with its predictions, it would be useful to perform the extensive quantum Monte Carlo calculations that would be needed to extend it to the thermodynamic limit.  The effects of temperature were also not explored in this study, although we expect them to be relatively weak when $T \ll T_{\lambda}$ where the bulk superfluid fraction is constant. 

In summary, our study demonstrates that alkali-metal pre-plated nanopores provide a promising, tunable environment for realizing compressible one-dimensional quantum liquids, extending the paradigm of surface wetting into the cylindrical geometry of nanoconfinement. This establishes a pathway toward experimental exploration of strongly correlated quantum hydrodynamics in a platform where geometry and interaction strength can be independently controlled.

\section{Data and Code Availability}
All code \cite{pimcrepo,relaxationrepo,paperrepo} and data \cite{datarepo} needed to reproduce the results of this study are available online.
\acknowledgements

S.~P, P.~S. and A.~D., acknowledge support from the U.S. Department of Energy, Office of Science, Office of Basic Energy Sciences, under Award Number DE-SC0024333.

\appendix

\section{Implementation details of the relaxation method in Sec.~\ref{sec:2}}
\label{app:ConDetails}

While the external potential $U_{\rm pore}$ experienced by $^4$He atoms has cylindrical symmetry, their interactions captured by $V$ do not.   Thus we solve the $N=2$ Schr\"odinger equation on a Cartesian grid which also allows us to efficiently compute the Laplacian operator using Fast Fourier Transform. We employ periodic boundary conditions in all three dimensions (for simplicity), although the helium atoms do not ``feel'' the box's $(x,y)$-boundaries due to the hard wall of the confinement potential.  Due to translational invariance along the $z$-axis, the six coordinates $(x_i,y_i,z_i)$, $i=1,2$ of the two helium atoms can be replaced with five: two explicit pairs $(x_1,y_1,x_2,y_2)$ and one relative coordinate $z=z_2-z_1$ 
corresponding to the inter-atomic separation along $z$. 
In these coordinates, the Laplacian operator becomes:
\begin{equation}
    \sum_{i=1}^2 \nabla_i^2 = \sum_{i=1}^2 
    \left( \frac{\partial^2}{\partial x_i^2} + \frac{\partial^2}{\partial y_i^2} \right) + 
    2 \frac{\partial^2}{\partial  z^2}.
\end{equation}
In our calculations, we solve for the wavefunction on a Cartesian grid $(x_1,y_1,x_2,y_2,z)$ with $72^4 \times 12$ points. The coarser resolution along $z$ is adequate due to (1) translational invariance, (2) the hardcore of the interaction (suppressing density at short $z$) and (3) the smooth behavior of $V$ for $r>r_m$. We have checked that discretization error is negligible by confirming convergence with additional calculations on a $96^4 \times 16$ grid.

\begin{figure}[h]
\begin{center}
   \includegraphics[width=\columnwidth]{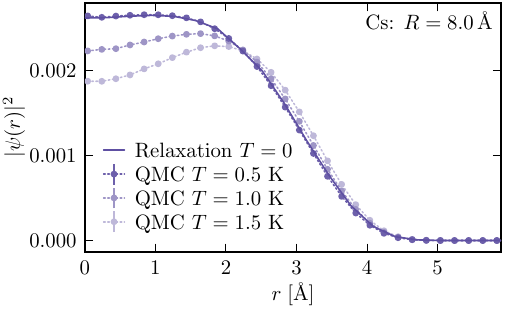}
\end{center}
\caption{The radial probability for $N=2$ $^4$He atoms confined inside a Cs pre-plated nanopore with $R=\SI{8.0}{\angstrom}$ and $L=\SI{25}{\angstrom}$ computed via the relaxation at $T=0$ on a $72^4 \times 12$ grid with the radial density computed via path integral quantum Monte Carlo at decreasing temperature. As the temperature decreases, we observe coincidence with the exact ground state solution within errorbars which are smaller than symbol size. Dashed lines are included as guides to the eye.} 
\label{fig:Pigs-compare}
\end{figure}

We consider that the iteration method described in Sec.~\ref{sec:2} is converged when $\sqrt{\langle \hat{L}_0 \psi_n| \hat{L}_0 \psi_n \rangle}$ decreases to $10^{-4}$, where
$\hat{L}_0 \psi_n$ is defined in Eq.~\eqref{eq:L0}. As a benchmark and test of the validity of our relaxation method we compare it to a $N = 2$ canonical quantum Monte Carlo simulation at decreasing temperatures as shown in Fig.~\ref{fig:Pigs-compare} and find exact agreement (within errorbars) at low temperature $T \le \SI{0.5}{\kelvin}$.

\bibliography{refs_nanopore}

\begin{thebibliography}{79}%
\makeatletter
\providecommand \@ifxundefined [1]{%
 \@ifx{#1\undefined}
}%
\providecommand \@ifnum [1]{%
 \ifnum #1\expandafter \@firstoftwo
 \else \expandafter \@secondoftwo
 \fi
}%
\providecommand \@ifx [1]{%
 \ifx #1\expandafter \@firstoftwo
 \else \expandafter \@secondoftwo
 \fi
}%
\providecommand \natexlab [1]{#1}%
\providecommand \enquote  [1]{``#1''}%
\providecommand \bibnamefont  [1]{#1}%
\providecommand \bibfnamefont [1]{#1}%
\providecommand \citenamefont [1]{#1}%
\providecommand \href@noop [0]{\@secondoftwo}%
\providecommand \href [0]{\begingroup \@sanitize@url \@href}%
\providecommand \@href[1]{\@@startlink{#1}\@@href}%
\providecommand \@@href[1]{\endgroup#1\@@endlink}%
\providecommand \@sanitize@url [0]{\catcode `\\12\catcode `\$12\catcode `\&12\catcode `\#12\catcode `\^12\catcode `\_12\catcode `\%12\relax}%
\providecommand \@@startlink[1]{}%
\providecommand \@@endlink[0]{}%
\providecommand \url  [0]{\begingroup\@sanitize@url \@url }%
\providecommand \@url [1]{\endgroup\@href {#1}{\urlprefix }}%
\providecommand \urlprefix  [0]{URL }%
\providecommand \Eprint [0]{\href }%
\providecommand \doibase [0]{https://doi.org/}%
\providecommand \selectlanguage [0]{\@gobble}%
\providecommand \bibinfo  [0]{\@secondoftwo}%
\providecommand \bibfield  [0]{\@secondoftwo}%
\providecommand \translation [1]{[#1]}%
\providecommand \BibitemOpen [0]{}%
\providecommand \bibitemStop [0]{}%
\providecommand \bibitemNoStop [0]{.\EOS\space}%
\providecommand \EOS [0]{\spacefactor3000\relax}%
\providecommand \BibitemShut  [1]{\csname bibitem#1\endcsname}%
\let\auto@bib@innerbib\@empty
\bibitem [{\citenamefont {Tomonaga}(1950)}]{Tomonaga:1950ak}%
  \BibitemOpen
  \bibfield  {author} {\bibinfo {author} {\bibfnamefont {S.~I.}\ \bibnamefont {Tomonaga}},\ }\bibfield  {title} {\bibinfo {title} {{R}emarks on {B}loch's {M}ethod of {S}ound {W}aves applied to {M}any-{F}ermion {P}roblems},\ }\href {https://doi.org/10.1143/ptp/5.4.544} {\bibfield  {journal} {\bibinfo  {journal} {Prog. Theor. Phys.}\ }\textbf {\bibinfo {volume} {5}},\ \bibinfo {pages} {544} (\bibinfo {year} {1950})}\BibitemShut {NoStop}%
\bibitem [{\citenamefont {Luttinger}(1963)}]{Luttinger:1963gd}%
  \BibitemOpen
  \bibfield  {author} {\bibinfo {author} {\bibfnamefont {J.~M.}\ \bibnamefont {Luttinger}},\ }\bibfield  {title} {\bibinfo {title} {{A}n {E}xactly {S}oluble {M}odel of a {M}any-{F}ermion {S}ystem},\ }\href {https://doi.org/10.1063/1.1704046} {\bibfield  {journal} {\bibinfo  {journal} {J. Math. Phys.}\ }\textbf {\bibinfo {volume} {4}},\ \bibinfo {pages} {1154} (\bibinfo {year} {1963})}\BibitemShut {NoStop}%
\bibitem [{\citenamefont {Mattis}\ and\ \citenamefont {Lieb}(1965)}]{Mattis:1965iq}%
  \BibitemOpen
  \bibfield  {author} {\bibinfo {author} {\bibfnamefont {D.~C.}\ \bibnamefont {Mattis}}\ and\ \bibinfo {author} {\bibfnamefont {E.~H.}\ \bibnamefont {Lieb}},\ }\bibfield  {title} {\bibinfo {title} {{E}xact {S}olution of a {M}any-{F}ermion {S}ystem and {I}ts {A}ssociated {B}oson {F}ield},\ }\href {https://doi.org/10.1063/1.1704281} {\bibfield  {journal} {\bibinfo  {journal} {J. Math. Phys.}\ }\textbf {\bibinfo {volume} {6}},\ \bibinfo {pages} {304} (\bibinfo {year} {1965})}\BibitemShut {NoStop}%
\bibitem [{\citenamefont {Haldane}(1981)}]{Haldane:1981eh}%
  \BibitemOpen
  \bibfield  {author} {\bibinfo {author} {\bibfnamefont {F.~D.~M.}\ \bibnamefont {Haldane}},\ }\bibfield  {title} {\bibinfo {title} {{Effective Harmonic-Fluid Approach to Low-Energy Properties of One-Dimensional Quantum Fluids}},\ }\href {https://doi.org/10.1103/PhysRevLett.47.1840} {\bibfield  {journal} {\bibinfo  {journal} {Phys. Rev. Lett.}\ }\textbf {\bibinfo {volume} {47}},\ \bibinfo {pages} {1840} (\bibinfo {year} {1981})}\BibitemShut {NoStop}%
\bibitem [{\citenamefont {Giamarchi}(2004)}]{Giamarchi:2004bk}%
  \BibitemOpen
  \bibfield  {author} {\bibinfo {author} {\bibfnamefont {T.}~\bibnamefont {Giamarchi}},\ }\href@noop {} {\emph {\bibinfo {title} {Quantum Physics in One Dimension}}}\ (\bibinfo  {publisher} {Clarendon Press},\ \bibinfo {address} {Oxford, U.K.},\ \bibinfo {year} {2004})\BibitemShut {NoStop}%
\bibitem [{\citenamefont {Lake}\ \emph {et~al.}(2005)\citenamefont {Lake}, \citenamefont {Tennant}, \citenamefont {Frost},\ and\ \citenamefont {Nagler}}]{Lake:2005ho}%
  \BibitemOpen
  \bibfield  {author} {\bibinfo {author} {\bibfnamefont {B.}~\bibnamefont {Lake}}, \bibinfo {author} {\bibfnamefont {D.~A.}\ \bibnamefont {Tennant}}, \bibinfo {author} {\bibfnamefont {C.~D.}\ \bibnamefont {Frost}},\ and\ \bibinfo {author} {\bibfnamefont {S.~E.}\ \bibnamefont {Nagler}},\ }\bibfield  {title} {\bibinfo {title} {{Q}uantum criticality and universal scaling of a quantum antiferromagnet},\ }\href {https://doi.org/10.1038/nmat1327} {\bibfield  {journal} {\bibinfo  {journal} {Nat. Mater.}\ }\textbf {\bibinfo {volume} {4}},\ \bibinfo {pages} {329} (\bibinfo {year} {2005})}\BibitemShut {NoStop}%
\bibitem [{\citenamefont {Jompol}\ \emph {et~al.}(2009)\citenamefont {Jompol}, \citenamefont {Ford}, \citenamefont {Griffiths}, \citenamefont {Farrer}, \citenamefont {Jones}, \citenamefont {Anderson}, \citenamefont {Ritchie}, \citenamefont {Silk},\ and\ \citenamefont {Schofield}}]{Jompol:2009sc}%
  \BibitemOpen
  \bibfield  {author} {\bibinfo {author} {\bibfnamefont {Y.}~\bibnamefont {Jompol}}, \bibinfo {author} {\bibfnamefont {C.~J.~B.}\ \bibnamefont {Ford}}, \bibinfo {author} {\bibfnamefont {J.~P.}\ \bibnamefont {Griffiths}}, \bibinfo {author} {\bibfnamefont {I.}~\bibnamefont {Farrer}}, \bibinfo {author} {\bibfnamefont {G.~A.~C.}\ \bibnamefont {Jones}}, \bibinfo {author} {\bibfnamefont {D.}~\bibnamefont {Anderson}}, \bibinfo {author} {\bibfnamefont {D.~A.}\ \bibnamefont {Ritchie}}, \bibinfo {author} {\bibfnamefont {T.~W.}\ \bibnamefont {Silk}},\ and\ \bibinfo {author} {\bibfnamefont {A.~J.}\ \bibnamefont {Schofield}},\ }\bibfield  {title} {\bibinfo {title} {Probing spin-charge separation in a tomonaga-luttinger liquid},\ }\href {https://doi.org/10.1126/science.1171769} {\bibfield  {journal} {\bibinfo  {journal} {Science}\ }\textbf {\bibinfo {volume} {325}},\ \bibinfo {pages} {597} (\bibinfo {year} {2009})}\BibitemShut {NoStop}%
\bibitem [{\citenamefont {Senaratne}\ \emph {et~al.}(2022)\citenamefont {Senaratne}, \citenamefont {Cavazos-Cavazos}, \citenamefont {Wang}, \citenamefont {He}, \citenamefont {Chang}, \citenamefont {Kafle}, \citenamefont {Pu}, \citenamefont {Guan},\ and\ \citenamefont {Hulet}}]{Senaratne:2022fg}%
  \BibitemOpen
  \bibfield  {author} {\bibinfo {author} {\bibfnamefont {R.}~\bibnamefont {Senaratne}}, \bibinfo {author} {\bibfnamefont {D.}~\bibnamefont {Cavazos-Cavazos}}, \bibinfo {author} {\bibfnamefont {S.}~\bibnamefont {Wang}}, \bibinfo {author} {\bibfnamefont {F.}~\bibnamefont {He}}, \bibinfo {author} {\bibfnamefont {Y.-T.}\ \bibnamefont {Chang}}, \bibinfo {author} {\bibfnamefont {A.}~\bibnamefont {Kafle}}, \bibinfo {author} {\bibfnamefont {H.}~\bibnamefont {Pu}}, \bibinfo {author} {\bibfnamefont {X.-W.}\ \bibnamefont {Guan}},\ and\ \bibinfo {author} {\bibfnamefont {R.~G.}\ \bibnamefont {Hulet}},\ }\bibfield  {title} {\bibinfo {title} {Spin-charge separation in a one-dimensional fermi gas with tunable interactions},\ }\href {https://doi.org/10.1126/science.abn1719} {\bibfield  {journal} {\bibinfo  {journal} {Science}\ }\textbf {\bibinfo {volume} {376}},\ \bibinfo {pages} {1305} (\bibinfo {year} {2022})}\BibitemShut {NoStop}%
\bibitem [{\citenamefont {Del~Maestro}\ \emph {et~al.}(2022)\citenamefont {Del~Maestro}, \citenamefont {Nichols}, \citenamefont {Prisk}, \citenamefont {Warren},\ and\ \citenamefont {Sokol}}]{DelMaestro22}%
  \BibitemOpen
  \bibfield  {author} {\bibinfo {author} {\bibfnamefont {A.}~\bibnamefont {Del~Maestro}}, \bibinfo {author} {\bibfnamefont {N.~S.}\ \bibnamefont {Nichols}}, \bibinfo {author} {\bibfnamefont {T.~R.}\ \bibnamefont {Prisk}}, \bibinfo {author} {\bibfnamefont {G.}~\bibnamefont {Warren}},\ and\ \bibinfo {author} {\bibfnamefont {P.~E.}\ \bibnamefont {Sokol}},\ }\bibfield  {title} {\bibinfo {title} {Experimental realization of one dimensional helium},\ }\href {https://doi.org/10.1038/s41467-022-30752-3} {\bibfield  {journal} {\bibinfo  {journal} {Nat. Commun.}\ }\textbf {\bibinfo {volume} {13}},\ \bibinfo {pages} {3168} (\bibinfo {year} {2022})}\BibitemShut {NoStop}%
\bibitem [{\citenamefont {Savard}\ \emph {et~al.}(2009)\citenamefont {Savard}, \citenamefont {Tremblay-Darveau},\ and\ \citenamefont {Gervais}}]{Savard:2009fc}%
  \BibitemOpen
  \bibfield  {author} {\bibinfo {author} {\bibfnamefont {M.}~\bibnamefont {Savard}}, \bibinfo {author} {\bibfnamefont {C.}~\bibnamefont {Tremblay-Darveau}},\ and\ \bibinfo {author} {\bibfnamefont {G.}~\bibnamefont {Gervais}},\ }\bibfield  {title} {\bibinfo {title} {Flow conductance of a single nanohole},\ }\href {https://doi.org/10.1103/PhysRevLett.103.104502} {\bibfield  {journal} {\bibinfo  {journal} {Phys. Rev. Lett.}\ }\textbf {\bibinfo {volume} {103}},\ \bibinfo {pages} {104502} (\bibinfo {year} {2009})}\BibitemShut {NoStop}%
\bibitem [{\citenamefont {Savard}\ \emph {et~al.}(2011)\citenamefont {Savard}, \citenamefont {Dauphinais},\ and\ \citenamefont {Gervais}}]{Savard:2011hd}%
  \BibitemOpen
  \bibfield  {author} {\bibinfo {author} {\bibfnamefont {M.}~\bibnamefont {Savard}}, \bibinfo {author} {\bibfnamefont {G.}~\bibnamefont {Dauphinais}},\ and\ \bibinfo {author} {\bibfnamefont {G.}~\bibnamefont {Gervais}},\ }\bibfield  {title} {\bibinfo {title} {Hydrodynamics of superfluid helium in a single nanohole},\ }\href {https://doi.org/10.1103/PhysRevLett.107.254501} {\bibfield  {journal} {\bibinfo  {journal} {Phys. Rev. Lett.}\ }\textbf {\bibinfo {volume} {107}},\ \bibinfo {pages} {254501} (\bibinfo {year} {2011})}\BibitemShut {NoStop}%
\bibitem [{\citenamefont {Velasco}\ \emph {et~al.}(2012)\citenamefont {Velasco}, \citenamefont {Friedman}, \citenamefont {Pevarnik}, \citenamefont {Siwy},\ and\ \citenamefont {Taborek}}]{Velasco:2012de}%
  \BibitemOpen
  \bibfield  {author} {\bibinfo {author} {\bibfnamefont {A.~E.}\ \bibnamefont {Velasco}}, \bibinfo {author} {\bibfnamefont {S.~G.}\ \bibnamefont {Friedman}}, \bibinfo {author} {\bibfnamefont {M.}~\bibnamefont {Pevarnik}}, \bibinfo {author} {\bibfnamefont {Z.~S.}\ \bibnamefont {Siwy}},\ and\ \bibinfo {author} {\bibfnamefont {P.}~\bibnamefont {Taborek}},\ }\bibfield  {title} {\bibinfo {title} {{Pressure-driven flow through a single nanopore}},\ }\href {https://doi.org/0.1103/PhysRevE.86.025302} {\bibfield  {journal} {\bibinfo  {journal} {Phys. Rev. E}\ }\textbf {\bibinfo {volume} {86}},\ \bibinfo {pages} {025302} (\bibinfo {year} {2012})}\BibitemShut {NoStop}%
\bibitem [{\citenamefont {Velasco}\ \emph {et~al.}(2014)\citenamefont {Velasco}, \citenamefont {Yang}, \citenamefont {Siwy}, \citenamefont {Toimil-Molares},\ and\ \citenamefont {Taborek}}]{Velasco:2014fe}%
  \BibitemOpen
  \bibfield  {author} {\bibinfo {author} {\bibfnamefont {A.~E.}\ \bibnamefont {Velasco}}, \bibinfo {author} {\bibfnamefont {C.}~\bibnamefont {Yang}}, \bibinfo {author} {\bibfnamefont {Z.~S.}\ \bibnamefont {Siwy}}, \bibinfo {author} {\bibfnamefont {M.~E.}\ \bibnamefont {Toimil-Molares}},\ and\ \bibinfo {author} {\bibfnamefont {P.}~\bibnamefont {Taborek}},\ }\bibfield  {title} {\bibinfo {title} {{Flow and evaporation in single micrometer and nanometer scale pipes}},\ }\href {https://doi.org/10.1063/1.4890985} {\bibfield  {journal} {\bibinfo  {journal} {App. Phys. Lett.}\ }\textbf {\bibinfo {volume} {105}},\ \bibinfo {pages} {033101} (\bibinfo {year} {2014})}\BibitemShut {NoStop}%
\bibitem [{\citenamefont {Duc}\ \emph {et~al.}(2015)\citenamefont {Duc}, \citenamefont {Savard}, \citenamefont {Petrescu}, \citenamefont {Rosenow}, \citenamefont {Del~Maestro},\ and\ \citenamefont {Gervais}}]{Duc:2015sa}%
  \BibitemOpen
  \bibfield  {author} {\bibinfo {author} {\bibfnamefont {P.-F.}\ \bibnamefont {Duc}}, \bibinfo {author} {\bibfnamefont {M.}~\bibnamefont {Savard}}, \bibinfo {author} {\bibfnamefont {M.}~\bibnamefont {Petrescu}}, \bibinfo {author} {\bibfnamefont {B.}~\bibnamefont {Rosenow}}, \bibinfo {author} {\bibfnamefont {A.}~\bibnamefont {Del~Maestro}},\ and\ \bibinfo {author} {\bibfnamefont {G.}~\bibnamefont {Gervais}},\ }\bibfield  {title} {\bibinfo {title} {{Critical flow and dissipation in a quasi-one-dimensional superfluid}},\ }\href {https://doi.org/10.1126/sciadv.1400222} {\bibfield  {journal} {\bibinfo  {journal} {Science Adv.}\ }\textbf {\bibinfo {volume} {1}},\ \bibinfo {pages} {e1400222} (\bibinfo {year} {2015})}\BibitemShut {NoStop}%
\bibitem [{\citenamefont {Botimer}\ and\ \citenamefont {Taborek}(2016)}]{Botimer:2016pd}%
  \BibitemOpen
  \bibfield  {author} {\bibinfo {author} {\bibfnamefont {J.}~\bibnamefont {Botimer}}\ and\ \bibinfo {author} {\bibfnamefont {P.}~\bibnamefont {Taborek}},\ }\bibfield  {title} {\bibinfo {title} {Pressure driven flow of superfluid $^{4}\mathrm{He}$ through a nanopipe},\ }\href {https://doi.org/10.1103/PhysRevFluids.1.054102} {\bibfield  {journal} {\bibinfo  {journal} {Phys. Rev. Fluids}\ }\textbf {\bibinfo {volume} {1}},\ \bibinfo {pages} {054102} (\bibinfo {year} {2016})}\BibitemShut {NoStop}%
\bibitem [{\citenamefont {Wada}\ \emph {et~al.}(2001)\citenamefont {Wada}, \citenamefont {Taniguchi}, \citenamefont {Ikegami}, \citenamefont {Inagaki},\ and\ \citenamefont {Fukushima}}]{Wada:2001hh}%
  \BibitemOpen
  \bibfield  {author} {\bibinfo {author} {\bibfnamefont {N.}~\bibnamefont {Wada}}, \bibinfo {author} {\bibfnamefont {J.}~\bibnamefont {Taniguchi}}, \bibinfo {author} {\bibfnamefont {H.}~\bibnamefont {Ikegami}}, \bibinfo {author} {\bibfnamefont {S.}~\bibnamefont {Inagaki}},\ and\ \bibinfo {author} {\bibfnamefont {Y.}~\bibnamefont {Fukushima}},\ }\bibfield  {title} {\bibinfo {title} {{Helium-4 Bose Fluids Formed in One-Dimensional 18 $\text{\AA}$ Diameter Pores}},\ }\href {https://doi.org/10.1103/PhysRevLett.86.4322} {\bibfield  {journal} {\bibinfo  {journal} {Phys. Rev. Lett.}\ }\textbf {\bibinfo {volume} {86}},\ \bibinfo {pages} {4322} (\bibinfo {year} {2001})}\BibitemShut {NoStop}%
\bibitem [{\citenamefont {Yamamoto}\ \emph {et~al.}(2004)\citenamefont {Yamamoto}, \citenamefont {Nakashima}, \citenamefont {Shibayama},\ and\ \citenamefont {Shirahama}}]{Yamamoto:2004ss}%
  \BibitemOpen
  \bibfield  {author} {\bibinfo {author} {\bibfnamefont {K.}~\bibnamefont {Yamamoto}}, \bibinfo {author} {\bibfnamefont {H.}~\bibnamefont {Nakashima}}, \bibinfo {author} {\bibfnamefont {Y.}~\bibnamefont {Shibayama}},\ and\ \bibinfo {author} {\bibfnamefont {K.}~\bibnamefont {Shirahama}},\ }\bibfield  {title} {\bibinfo {title} {Anomalous suppression of superfluidity in $^{4}\mathrm{H}\mathrm{e}$ confined in a nanoporous glass},\ }\href {https://doi.org/10.1103/PhysRevLett.93.075302} {\bibfield  {journal} {\bibinfo  {journal} {Phys. Rev. Lett.}\ }\textbf {\bibinfo {volume} {93}},\ \bibinfo {pages} {075302} (\bibinfo {year} {2004})}\BibitemShut {NoStop}%
\bibitem [{\citenamefont {Taniguchi}\ \emph {et~al.}(2005)\citenamefont {Taniguchi}, \citenamefont {Yamaguchi}, \citenamefont {Ishimoto}, \citenamefont {Ikegami}, \citenamefont {Matsushita}, \citenamefont {Wada}, \citenamefont {Gatica}, \citenamefont {Cole}, \citenamefont {Ancilotto}, \citenamefont {Inagaki},\ and\ \citenamefont {Fukushima}}]{Taniguchi:2005pr}%
  \BibitemOpen
  \bibfield  {author} {\bibinfo {author} {\bibfnamefont {J.}~\bibnamefont {Taniguchi}}, \bibinfo {author} {\bibfnamefont {A.}~\bibnamefont {Yamaguchi}}, \bibinfo {author} {\bibfnamefont {H.}~\bibnamefont {Ishimoto}}, \bibinfo {author} {\bibfnamefont {H.}~\bibnamefont {Ikegami}}, \bibinfo {author} {\bibfnamefont {T.}~\bibnamefont {Matsushita}}, \bibinfo {author} {\bibfnamefont {N.}~\bibnamefont {Wada}}, \bibinfo {author} {\bibfnamefont {S.~M.}\ \bibnamefont {Gatica}}, \bibinfo {author} {\bibfnamefont {M.~W.}\ \bibnamefont {Cole}}, \bibinfo {author} {\bibfnamefont {F.}~\bibnamefont {Ancilotto}}, \bibinfo {author} {\bibfnamefont {S.}~\bibnamefont {Inagaki}},\ and\ \bibinfo {author} {\bibfnamefont {Y.}~\bibnamefont {Fukushima}},\ }\bibfield  {title} {\bibinfo {title} {Possible one-dimensional $^{3}\mathrm{H}\mathrm{e}$ quantum fluid formed in nanopores},\ }\href {https://doi.org/10.1103/PhysRevLett.94.065301} {\bibfield  {journal} {\bibinfo  {journal} {Phys. Rev. Lett.}\ }\textbf {\bibinfo {volume} {94}},\ \bibinfo {pages} {065301} (\bibinfo {year} {2005})}\BibitemShut {NoStop}%
\bibitem [{\citenamefont {Toda}\ \emph {et~al.}(2007)\citenamefont {Toda}, \citenamefont {Hieda}, \citenamefont {Matsushita}, \citenamefont {Wada}, \citenamefont {Taniguchi}, \citenamefont {Ikegami}, \citenamefont {Inagaki},\ and\ \citenamefont {Fukushima}}]{Toda:2007sf}%
  \BibitemOpen
  \bibfield  {author} {\bibinfo {author} {\bibfnamefont {R.}~\bibnamefont {Toda}}, \bibinfo {author} {\bibfnamefont {M.}~\bibnamefont {Hieda}}, \bibinfo {author} {\bibfnamefont {T.}~\bibnamefont {Matsushita}}, \bibinfo {author} {\bibfnamefont {N.}~\bibnamefont {Wada}}, \bibinfo {author} {\bibfnamefont {J.}~\bibnamefont {Taniguchi}}, \bibinfo {author} {\bibfnamefont {H.}~\bibnamefont {Ikegami}}, \bibinfo {author} {\bibfnamefont {S.}~\bibnamefont {Inagaki}},\ and\ \bibinfo {author} {\bibfnamefont {Y.}~\bibnamefont {Fukushima}},\ }\bibfield  {title} {\bibinfo {title} {{Superfluidity of He4 in One and Three Dimensions Realized in Nanopores}},\ }\href {https://doi.org/10.1103/PhysRevLett.99.255301} {\bibfield  {journal} {\bibinfo  {journal} {Phys. Rev. Lett.}\ }\textbf {\bibinfo {volume} {99}},\ \bibinfo {pages} {255301} (\bibinfo {year} {2007})}\BibitemShut {NoStop}%
\bibitem [{\citenamefont {Ikegami}\ \emph {et~al.}(2005)\citenamefont {Ikegami}, \citenamefont {Yamato}, \citenamefont {Okuno}, \citenamefont {Taniguchi},\ and\ \citenamefont {Wada}}]{Ikegami2005}%
  \BibitemOpen
  \bibfield  {author} {\bibinfo {author} {\bibfnamefont {H.}~\bibnamefont {Ikegami}}, \bibinfo {author} {\bibfnamefont {Y.}~\bibnamefont {Yamato}}, \bibinfo {author} {\bibfnamefont {T.}~\bibnamefont {Okuno}}, \bibinfo {author} {\bibfnamefont {J.}~\bibnamefont {Taniguchi}},\ and\ \bibinfo {author} {\bibfnamefont {N.}~\bibnamefont {Wada}},\ }\bibfield  {title} {\bibinfo {title} {Observation of 4he superfluidity in 1.8 nm-pores},\ }\href {https://doi.org/10.1007/s10909-005-1546-2} {\bibfield  {journal} {\bibinfo  {journal} {J. Low Temp. Phys.}\ }\textbf {\bibinfo {volume} {138}},\ \bibinfo {pages} {171} (\bibinfo {year} {2005})}\BibitemShut {NoStop}%
\bibitem [{\citenamefont {Azuah}\ \emph {et~al.}(2013)\citenamefont {Azuah}, \citenamefont {Diallo}, \citenamefont {Adams}, \citenamefont {Kirichek},\ and\ \citenamefont {Glyde}}]{Azuah2013}%
  \BibitemOpen
  \bibfield  {author} {\bibinfo {author} {\bibfnamefont {R.~T.}\ \bibnamefont {Azuah}}, \bibinfo {author} {\bibfnamefont {S.~O.}\ \bibnamefont {Diallo}}, \bibinfo {author} {\bibfnamefont {M.~A.}\ \bibnamefont {Adams}}, \bibinfo {author} {\bibfnamefont {O.}~\bibnamefont {Kirichek}},\ and\ \bibinfo {author} {\bibfnamefont {H.~R.}\ \bibnamefont {Glyde}},\ }\bibfield  {title} {\bibinfo {title} {{Phonon-roton modes of liquid ${}^{4}$He beyond the roton in the porous medium MCM-41}},\ }\href {https://doi.org/10.1103/PhysRevB.88.024510} {\bibfield  {journal} {\bibinfo  {journal} {Phys. Rev. B}\ }\textbf {\bibinfo {volume} {88}},\ \bibinfo {pages} {024510} (\bibinfo {year} {2013})}\BibitemShut {NoStop}%
\bibitem [{\citenamefont {Prisk}\ \emph {et~al.}(2013)\citenamefont {Prisk}, \citenamefont {Das}, \citenamefont {Diallo}, \citenamefont {Ehlers}, \citenamefont {Podlesnyak}, \citenamefont {Wada}, \citenamefont {Inagaki},\ and\ \citenamefont {Sokol}}]{Prisk:2013br}%
  \BibitemOpen
  \bibfield  {author} {\bibinfo {author} {\bibfnamefont {T.~R.}\ \bibnamefont {Prisk}}, \bibinfo {author} {\bibfnamefont {N.~C.}\ \bibnamefont {Das}}, \bibinfo {author} {\bibfnamefont {S.~O.}\ \bibnamefont {Diallo}}, \bibinfo {author} {\bibfnamefont {G.}~\bibnamefont {Ehlers}}, \bibinfo {author} {\bibfnamefont {A.~A.}\ \bibnamefont {Podlesnyak}}, \bibinfo {author} {\bibfnamefont {N.}~\bibnamefont {Wada}}, \bibinfo {author} {\bibfnamefont {S.}~\bibnamefont {Inagaki}},\ and\ \bibinfo {author} {\bibfnamefont {P.~E.}\ \bibnamefont {Sokol}},\ }\bibfield  {title} {\bibinfo {title} {{P}hases of superfluid helium in smooth cylindrical pores},\ }\href {https://doi.org/10.1103/physrevb.88.014521} {\bibfield  {journal} {\bibinfo  {journal} {Phys. Rev. B}\ }\textbf {\bibinfo {volume} {88}},\ \bibinfo {pages} {014521} (\bibinfo {year} {2013})}\BibitemShut {NoStop}%
\bibitem [{\citenamefont {Yager}\ \emph {et~al.}(2013)\citenamefont {Yager}, \citenamefont {Ny{\'e}ki}, \citenamefont {Casey}, \citenamefont {Cowan}, \citenamefont {Lusher},\ and\ \citenamefont {Saunders}}]{Yager:2013cva}%
  \BibitemOpen
  \bibfield  {author} {\bibinfo {author} {\bibfnamefont {B.}~\bibnamefont {Yager}}, \bibinfo {author} {\bibfnamefont {J.}~\bibnamefont {Ny{\'e}ki}}, \bibinfo {author} {\bibfnamefont {A.}~\bibnamefont {Casey}}, \bibinfo {author} {\bibfnamefont {B.~P.}\ \bibnamefont {Cowan}}, \bibinfo {author} {\bibfnamefont {C.~P.}\ \bibnamefont {Lusher}},\ and\ \bibinfo {author} {\bibfnamefont {J.}~\bibnamefont {Saunders}},\ }\bibfield  {title} {\bibinfo {title} {{NMR Signature of One-Dimensional Behavior of 3He in Nanopores}},\ }\href {https://doi.org/10.1103/PhysRevLett.111.215303} {\bibfield  {journal} {\bibinfo  {journal} {Phys. Rev. Lett.}\ }\textbf {\bibinfo {volume} {111}},\ \bibinfo {pages} {215303} (\bibinfo {year} {2013})}\BibitemShut {NoStop}%
\bibitem [{\citenamefont {Demura}\ \emph {et~al.}(2015)\citenamefont {Demura}, \citenamefont {Taniguchi},\ and\ \citenamefont {Suzuki}}]{Demura:2015hq}%
  \BibitemOpen
  \bibfield  {author} {\bibinfo {author} {\bibfnamefont {K.}~\bibnamefont {Demura}}, \bibinfo {author} {\bibfnamefont {J.}~\bibnamefont {Taniguchi}},\ and\ \bibinfo {author} {\bibfnamefont {M.}~\bibnamefont {Suzuki}},\ }\bibfield  {title} {\bibinfo {title} {{Dynamical Superfluid Response of 3He-4He Solutions Confined in a Nanometer-Size Channel}},\ }\href {https://doi.org/10.7566/JPSJ.84.094604} {\bibfield  {journal} {\bibinfo  {journal} {J. Phys. Soc. Jpn.}\ }\textbf {\bibinfo {volume} {84}},\ \bibinfo {pages} {094604} (\bibinfo {year} {2015})}\BibitemShut {NoStop}%
\bibitem [{\citenamefont {Demura}\ \emph {et~al.}(2017)\citenamefont {Demura}, \citenamefont {Taniguchi},\ and\ \citenamefont {Suzuki}}]{Demura:2017zj}%
  \BibitemOpen
  \bibfield  {author} {\bibinfo {author} {\bibfnamefont {K.}~\bibnamefont {Demura}}, \bibinfo {author} {\bibfnamefont {J.}~\bibnamefont {Taniguchi}},\ and\ \bibinfo {author} {\bibfnamefont {M.}~\bibnamefont {Suzuki}},\ }\bibfield  {title} {\bibinfo {title} {{T}wofold {T}orsional {O}scillator {E}xperiments from {F}ilm to {P}ressurized {L}iquid 4{H}e in a {N}anometer-{S}ize {C}hannel},\ }\href {https://doi.org/10.7566/jpsj.86.114601} {\bibfield  {journal} {\bibinfo  {journal} {J. Phys. Soc. Jpn.}\ }\textbf {\bibinfo {volume} {86}},\ \bibinfo {pages} {114601} (\bibinfo {year} {2017})}\BibitemShut {NoStop}%
\bibitem [{\citenamefont {Bryan}\ \emph {et~al.}(2017)\citenamefont {Bryan}, \citenamefont {Prisk}, \citenamefont {Sherline}, \citenamefont {Diallo},\ and\ \citenamefont {Sokol}}]{Bryan:2017gz}%
  \BibitemOpen
  \bibfield  {author} {\bibinfo {author} {\bibfnamefont {M.~S.}\ \bibnamefont {Bryan}}, \bibinfo {author} {\bibfnamefont {T.~R.}\ \bibnamefont {Prisk}}, \bibinfo {author} {\bibfnamefont {T.~E.}\ \bibnamefont {Sherline}}, \bibinfo {author} {\bibfnamefont {S.~O.}\ \bibnamefont {Diallo}},\ and\ \bibinfo {author} {\bibfnamefont {P.~E.}\ \bibnamefont {Sokol}},\ }\bibfield  {title} {\bibinfo {title} {{B}ulklike excitations in nanoconfined liquid helium},\ }\href {https://doi.org/10.1103/physrevb.95.144509} {\bibfield  {journal} {\bibinfo  {journal} {Phys. Rev. B}\ }\textbf {\bibinfo {volume} {95}},\ \bibinfo {pages} {144509} (\bibinfo {year} {2017})}\BibitemShut {NoStop}%
\bibitem [{\citenamefont {Taniguchi}\ \emph {et~al.}(2018)\citenamefont {Taniguchi}, \citenamefont {Taniguchi},\ and\ \citenamefont {Suzuki}}]{Taniguchi_2018}%
  \BibitemOpen
  \bibfield  {author} {\bibinfo {author} {\bibfnamefont {K.}~\bibnamefont {Taniguchi}}, \bibinfo {author} {\bibfnamefont {J.}~\bibnamefont {Taniguchi}},\ and\ \bibinfo {author} {\bibfnamefont {M.}~\bibnamefont {Suzuki}},\ }\bibfield  {title} {\bibinfo {title} {Torsional oscillator measurements of liquid 4he confined in 2.5-nm channel of fsm},\ }\href {https://doi.org/10.1088/1742-6596/969/1/012005} {\bibfield  {journal} {\bibinfo  {journal} {J. Phys. Conf. Ser.}\ }\textbf {\bibinfo {volume} {969}},\ \bibinfo {pages} {012005} (\bibinfo {year} {2018})}\BibitemShut {NoStop}%
\bibitem [{\citenamefont {Bossy}\ \emph {et~al.}(2019)\citenamefont {Bossy}, \citenamefont {Ollivier},\ and\ \citenamefont {Glyde}}]{Bossy:2019}%
  \BibitemOpen
  \bibfield  {author} {\bibinfo {author} {\bibfnamefont {J.}~\bibnamefont {Bossy}}, \bibinfo {author} {\bibfnamefont {J.}~\bibnamefont {Ollivier}},\ and\ \bibinfo {author} {\bibfnamefont {H.~R.}\ \bibnamefont {Glyde}},\ }\bibfield  {title} {\bibinfo {title} {Phonons, rotons, and localized bose-einstein condensation in liquid $^{4}\mathrm{He}$ confined in nanoporous fsm-16},\ }\href {https://doi.org/10.1103/PhysRevB.99.165425} {\bibfield  {journal} {\bibinfo  {journal} {Phys. Rev. B}\ }\textbf {\bibinfo {volume} {99}},\ \bibinfo {pages} {165425} (\bibinfo {year} {2019})}\BibitemShut {NoStop}%
\bibitem [{\citenamefont {Taniguchi}\ \emph {et~al.}(2020)\citenamefont {Taniguchi}, \citenamefont {Taniguchi}, \citenamefont {Kanno},\ and\ \citenamefont {Suzuki}}]{Taniguchi2020}%
  \BibitemOpen
  \bibfield  {author} {\bibinfo {author} {\bibfnamefont {J.}~\bibnamefont {Taniguchi}}, \bibinfo {author} {\bibfnamefont {K.}~\bibnamefont {Taniguchi}}, \bibinfo {author} {\bibfnamefont {K.}~\bibnamefont {Kanno}},\ and\ \bibinfo {author} {\bibfnamefont {M.}~\bibnamefont {Suzuki}},\ }\bibfield  {title} {\bibinfo {title} {Possible thermodynamical phase slips in superfluid 4he confined in a 2.5-nm channel of fsm},\ }\href {https://doi.org/10.1007/s10909-020-02355-z} {\bibfield  {journal} {\bibinfo  {journal} {J. Low Temp. Phys.}\ }\textbf {\bibinfo {volume} {201}},\ \bibinfo {pages} {139–145} (\bibinfo {year} {2020})}\BibitemShut {NoStop}%
\bibitem [{\citenamefont {Kuribara}\ \emph {et~al.}(2023)\citenamefont {Kuribara}, \citenamefont {Taniguchi},\ and\ \citenamefont {Suzuki}}]{Kuribara:2023sf}%
  \BibitemOpen
  \bibfield  {author} {\bibinfo {author} {\bibfnamefont {M.}~\bibnamefont {Kuribara}}, \bibinfo {author} {\bibfnamefont {J.}~\bibnamefont {Taniguchi}},\ and\ \bibinfo {author} {\bibfnamefont {M.}~\bibnamefont {Suzuki}},\ }\bibfield  {title} {\bibinfo {title} {{Superflow of 4He through an Oriented Nanometer-sized Porous Membrane}},\ }\href {https://doi.org/10.7566/jpscp.38.011006} {\bibfield  {journal} {\bibinfo  {journal} {J. Phys. Soc. Jap.}\ }\textbf {\bibinfo {volume} {38}},\ \bibinfo {pages} {011006} (\bibinfo {year} {2023})}\BibitemShut {NoStop}%
\bibitem [{\citenamefont {Galli}\ \emph {et~al.}(2014)\citenamefont {Galli}, \citenamefont {Reatto},\ and\ \citenamefont {Rossi}}]{Galli:2014uw}%
  \BibitemOpen
  \bibfield  {author} {\bibinfo {author} {\bibfnamefont {D.~E.}\ \bibnamefont {Galli}}, \bibinfo {author} {\bibfnamefont {L.}~\bibnamefont {Reatto}},\ and\ \bibinfo {author} {\bibfnamefont {M.}~\bibnamefont {Rossi}},\ }\bibfield  {title} {\bibinfo {title} {{Q}uantum {M}onte {C}arlo study of a vortex in superfluid 4{H}e and search for a vortex state in the solid},\ }\href {https://doi.org/10.1103/physrevb.89.224516} {\bibfield  {journal} {\bibinfo  {journal} {Phys. Rev. B}\ }\textbf {\bibinfo {volume} {89}},\ \bibinfo {pages} {224516} (\bibinfo {year} {2014})}\BibitemShut {NoStop}%
\bibitem [{\citenamefont {Amelio}\ \emph {et~al.}(2018)\citenamefont {Amelio}, \citenamefont {Galli},\ and\ \citenamefont {Reatto}}]{Amelio:2018cx}%
  \BibitemOpen
  \bibfield  {author} {\bibinfo {author} {\bibfnamefont {I.}~\bibnamefont {Amelio}}, \bibinfo {author} {\bibfnamefont {D.~E.}\ \bibnamefont {Galli}},\ and\ \bibinfo {author} {\bibfnamefont {L.}~\bibnamefont {Reatto}},\ }\bibfield  {title} {\bibinfo {title} {{P}robing {Q}uantum {T}urbulence in {H}e4},\ }\href {https://doi.org/10.1103/physrevlett.121.015302} {\bibfield  {journal} {\bibinfo  {journal} {Phys. Rev. Lett.}\ }\textbf {\bibinfo {volume} {121}},\ \bibinfo {pages} {015302} (\bibinfo {year} {2018})}\BibitemShut {NoStop}%
\bibitem [{\citenamefont {Chakravarty}(1997)}]{Chakravarty:1997eq}%
  \BibitemOpen
  \bibfield  {author} {\bibinfo {author} {\bibfnamefont {C.}~\bibnamefont {Chakravarty}},\ }\bibfield  {title} {\bibinfo {title} {{Q}uantum {A}dsorbates: {P}ath {I}ntegral {M}onte {C}arlo {S}imulations of {H}elium in {S}ilicalite},\ }\href {https://doi.org/10.1021/jp962155s} {\bibfield  {journal} {\bibinfo  {journal} {J. Phys. Chem. B}\ }\textbf {\bibinfo {volume} {101}},\ \bibinfo {pages} {1878} (\bibinfo {year} {1997})}\BibitemShut {NoStop}%
\bibitem [{\citenamefont {Cole}\ \emph {et~al.}(2000)\citenamefont {Cole}, \citenamefont {Crespi}, \citenamefont {Stan}, \citenamefont {Ebner}, \citenamefont {Hartman}, \citenamefont {Moroni},\ and\ \citenamefont {Boninsegni}}]{Cole:2000tz}%
  \BibitemOpen
  \bibfield  {author} {\bibinfo {author} {\bibfnamefont {M.~W.}\ \bibnamefont {Cole}}, \bibinfo {author} {\bibfnamefont {V.~H.}\ \bibnamefont {Crespi}}, \bibinfo {author} {\bibfnamefont {G.}~\bibnamefont {Stan}}, \bibinfo {author} {\bibfnamefont {C.}~\bibnamefont {Ebner}}, \bibinfo {author} {\bibfnamefont {J.~M.}\ \bibnamefont {Hartman}}, \bibinfo {author} {\bibfnamefont {S.}~\bibnamefont {Moroni}},\ and\ \bibinfo {author} {\bibfnamefont {M.}~\bibnamefont {Boninsegni}},\ }\bibfield  {title} {\bibinfo {title} {{C}ondensation of {H}elium in {N}anotube {B}undles},\ }\href {https://doi.org/10.1103/physrevlett.84.3883} {\bibfield  {journal} {\bibinfo  {journal} {Phys. Rev. Lett.}\ }\textbf {\bibinfo {volume} {84}},\ \bibinfo {pages} {3883} (\bibinfo {year} {2000})}\BibitemShut {NoStop}%
\bibitem [{\citenamefont {Gordillo}\ \emph {et~al.}(2000)\citenamefont {Gordillo}, \citenamefont {Boronat},\ and\ \citenamefont {Casulleras}}]{Gordillo:2000qo}%
  \BibitemOpen
  \bibfield  {author} {\bibinfo {author} {\bibfnamefont {M.~C.}\ \bibnamefont {Gordillo}}, \bibinfo {author} {\bibfnamefont {J.}~\bibnamefont {Boronat}},\ and\ \bibinfo {author} {\bibfnamefont {J.}~\bibnamefont {Casulleras}},\ }\bibfield  {title} {\bibinfo {title} {Quasi-one-dimensional ${}^{4}\mathrm{He}$ inside carbon nanotubes},\ }\href {https://doi.org/10.1103/PhysRevB.61.R878} {\bibfield  {journal} {\bibinfo  {journal} {Phys. Rev. B}\ }\textbf {\bibinfo {volume} {61}},\ \bibinfo {pages} {R878} (\bibinfo {year} {2000})}\BibitemShut {NoStop}%
\bibitem [{\citenamefont {Rossi}\ \emph {et~al.}(2006)\citenamefont {Rossi}, \citenamefont {Galli},\ and\ \citenamefont {Reatto}}]{Rossi:2006cg}%
  \BibitemOpen
  \bibfield  {author} {\bibinfo {author} {\bibfnamefont {M.}~\bibnamefont {Rossi}}, \bibinfo {author} {\bibfnamefont {D.~E.}\ \bibnamefont {Galli}},\ and\ \bibinfo {author} {\bibfnamefont {L.}~\bibnamefont {Reatto}},\ }\bibfield  {title} {\bibinfo {title} {{P}ressurized 4{H}e in {C}ylindrical and in {H}exagonal {P}ores},\ }\href {https://doi.org/10.1007/s10909-006-9265-x} {\bibfield  {journal} {\bibinfo  {journal} {J. Low Temp. Phys.}\ }\textbf {\bibinfo {volume} {146}},\ \bibinfo {pages} {95} (\bibinfo {year} {2006})}\BibitemShut {NoStop}%
\bibitem [{\citenamefont {Del~Maestro}\ \emph {et~al.}(2011)\citenamefont {Del~Maestro}, \citenamefont {Boninsegni},\ and\ \citenamefont {Affleck}}]{DelMaestro:2011et}%
  \BibitemOpen
  \bibfield  {author} {\bibinfo {author} {\bibfnamefont {A.}~\bibnamefont {Del~Maestro}}, \bibinfo {author} {\bibfnamefont {M.}~\bibnamefont {Boninsegni}},\ and\ \bibinfo {author} {\bibfnamefont {I.}~\bibnamefont {Affleck}},\ }\bibfield  {title} {\bibinfo {title} {$^4${He} {L}uttinger {L}iquid in {N}anopores},\ }\href {https://doi.org/10.1103/physrevlett.106.105303} {\bibfield  {journal} {\bibinfo  {journal} {Phys. Rev. Lett.}\ }\textbf {\bibinfo {volume} {106}},\ \bibinfo {pages} {105303} (\bibinfo {year} {2011})}\BibitemShut {NoStop}%
\bibitem [{\citenamefont {{Del Maestro}}(2012)}]{DelMaestro:2012td}%
  \BibitemOpen
  \bibfield  {author} {\bibinfo {author} {\bibfnamefont {A.}~\bibnamefont {{Del Maestro}}},\ }\bibfield  {title} {\bibinfo {title} {{ A Luttinger Liquid Core Inside Helium-4 Filled Nanopores}},\ }\href {https://doi.org/10.1142/s021797921244002x} {\bibfield  {journal} {\bibinfo  {journal} {Int. J. Mod. Phys. B}\ }\textbf {\bibinfo {volume} {26}},\ \bibinfo {pages} {1244002} (\bibinfo {year} {2012})}\BibitemShut {NoStop}%
\bibitem [{\citenamefont {Kulchytskyy}\ \emph {et~al.}(2013)\citenamefont {Kulchytskyy}, \citenamefont {Gervais},\ and\ \citenamefont {{Del Maestro}}}]{Kulchytskyy:2013qq}%
  \BibitemOpen
  \bibfield  {author} {\bibinfo {author} {\bibfnamefont {B.}~\bibnamefont {Kulchytskyy}}, \bibinfo {author} {\bibfnamefont {G.}~\bibnamefont {Gervais}},\ and\ \bibinfo {author} {\bibfnamefont {A.}~\bibnamefont {{Del Maestro}}},\ }\bibfield  {title} {\bibinfo {title} {{L}ocal superfluidity at the nanoscale},\ }\href {https://doi.org/10.1103/physrevb.88.064512} {\bibfield  {journal} {\bibinfo  {journal} {Phys. Rev. B}\ }\textbf {\bibinfo {volume} {88}},\ \bibinfo {pages} {064512} (\bibinfo {year} {2013})}\BibitemShut {NoStop}%
\bibitem [{\citenamefont {Marki{\'c}}(2015)}]{Markic:2015bu}%
  \BibitemOpen
  \bibfield  {author} {\bibinfo {author} {\bibfnamefont {H.~R.}\ \bibnamefont {Marki{\'c}}, \bibfnamefont {L.~Vranje{\v s}and~Glyde}},\ }\bibfield  {title} {\bibinfo {title} {Superfluidity, {BEC}, and dimensions of liquid $^{4}\mathrm{He}$ in nanopores},\ }\href {https://link.aps.org/doi/10.1103/PhysRevB.92.064510} {\bibfield  {journal} {\bibinfo  {journal} {Phys. Rev. B}\ }\textbf {\bibinfo {volume} {92}},\ \bibinfo {pages} {064510} (\bibinfo {year} {2015})}\BibitemShut {NoStop}%
\bibitem [{\citenamefont {Marki{\'c}}\ \emph {et~al.}(2018)\citenamefont {Marki{\'c}}, \citenamefont {Vrcan}, \citenamefont {Zuhrianda},\ and\ \citenamefont {Glyde}}]{Markic:2018bw}%
  \BibitemOpen
  \bibfield  {author} {\bibinfo {author} {\bibfnamefont {L.~V.}\ \bibnamefont {Marki{\'c}}}, \bibinfo {author} {\bibfnamefont {H.}~\bibnamefont {Vrcan}}, \bibinfo {author} {\bibfnamefont {Z.}~\bibnamefont {Zuhrianda}},\ and\ \bibinfo {author} {\bibfnamefont {H.~R.}\ \bibnamefont {Glyde}},\ }\bibfield  {title} {\bibinfo {title} {{S}uperfluidity, {B}ose-{E}instein condensation, and structure in one-dimensional {L}uttinger liquids},\ }\href {https://doi.org/10.1103/PhysRevB.97.014513} {\bibfield  {journal} {\bibinfo  {journal} {Phys. Rev. B}\ }\textbf {\bibinfo {volume} {97}},\ \bibinfo {pages} {014513} (\bibinfo {year} {2018})}\BibitemShut {NoStop}%
\bibitem [{\citenamefont {Marki{\' c}}\ \emph {et~al.}(2020)\citenamefont {Marki{\' c}}, \citenamefont {D{\v z}elalija},\ and\ \citenamefont {Glyde}}]{Markic:2020}%
  \BibitemOpen
  \bibfield  {author} {\bibinfo {author} {\bibfnamefont {L.~V.}\ \bibnamefont {Marki{\' c}}}, \bibinfo {author} {\bibfnamefont {K.}~\bibnamefont {D{\v z}elalija}},\ and\ \bibinfo {author} {\bibfnamefont {H.~R.}\ \bibnamefont {Glyde}},\ }\bibfield  {title} {\bibinfo {title} {{C}rossover from one to two dimensions in liquid $^4${H}e in a nanopore},\ }\href {https://doi.org/10.1103/physrevb.101.104505} {\bibfield  {journal} {\bibinfo  {journal} {Phys. Rev. B}\ }\textbf {\bibinfo {volume} {101}},\ \bibinfo {pages} {104505} (\bibinfo {year} {2020})}\BibitemShut {NoStop}%
\bibitem [{\citenamefont {Nava}\ \emph {et~al.}(2022)\citenamefont {Nava}, \citenamefont {Giuliano}, \citenamefont {Nguyen},\ and\ \citenamefont {Boninsegni}}]{Nava:2022hk}%
  \BibitemOpen
  \bibfield  {author} {\bibinfo {author} {\bibfnamefont {A.}~\bibnamefont {Nava}}, \bibinfo {author} {\bibfnamefont {D.}~\bibnamefont {Giuliano}}, \bibinfo {author} {\bibfnamefont {P.~H.}\ \bibnamefont {Nguyen}},\ and\ \bibinfo {author} {\bibfnamefont {M.}~\bibnamefont {Boninsegni}},\ }\bibfield  {title} {\bibinfo {title} {{Q}uasi-one-dimensional 4{H}e in nanopores},\ }\href {https://doi.org/10.1103/physrevb.105.085402} {\bibfield  {journal} {\bibinfo  {journal} {Phys. Rev. B}\ }\textbf {\bibinfo {volume} {105}},\ \bibinfo {pages} {085402} (\bibinfo {year} {2022})}\BibitemShut {NoStop}%
\bibitem [{\citenamefont {Kresge}\ \emph {et~al.}(1992)\citenamefont {Kresge}, \citenamefont {Leonowicz}, \citenamefont {Roth}, \citenamefont {Vartuli},\ and\ \citenamefont {Beck}}]{Kresge:1992vv}%
  \BibitemOpen
  \bibfield  {author} {\bibinfo {author} {\bibfnamefont {C.~T.}\ \bibnamefont {Kresge}}, \bibinfo {author} {\bibfnamefont {M.~E.}\ \bibnamefont {Leonowicz}}, \bibinfo {author} {\bibfnamefont {W.~J.}\ \bibnamefont {Roth}}, \bibinfo {author} {\bibfnamefont {J.~C.}\ \bibnamefont {Vartuli}},\ and\ \bibinfo {author} {\bibfnamefont {J.~S.}\ \bibnamefont {Beck}},\ }\bibfield  {title} {\bibinfo {title} {{O}rdered mesoporous molecular sieves synthesized by a liquid-crystal template mechanism},\ }\href {https://doi.org/10.1038/359710a0} {\bibfield  {journal} {\bibinfo  {journal} {Nature}\ }\textbf {\bibinfo {volume} {359}},\ \bibinfo {pages} {710} (\bibinfo {year} {1992})}\BibitemShut {NoStop}%
\bibitem [{\citenamefont {Sonwane}\ and\ \citenamefont {Bhatia}(1999)}]{Sonwane:1999xl}%
  \BibitemOpen
  \bibfield  {author} {\bibinfo {author} {\bibfnamefont {C.~G.}\ \bibnamefont {Sonwane}}\ and\ \bibinfo {author} {\bibfnamefont {S.~K.}\ \bibnamefont {Bhatia}},\ }\bibfield  {title} {\bibinfo {title} {{S}tructural {C}haracterization of {MCM-41} over a {W}ide {R}ange of {L}ength {S}cales},\ }\href {https://doi.org/10.1021/la9807614} {\bibfield  {journal} {\bibinfo  {journal} {Langmuir}\ }\textbf {\bibinfo {volume} {15}},\ \bibinfo {pages} {2809} (\bibinfo {year} {1999})}\BibitemShut {NoStop}%
\bibitem [{\citenamefont {Inagaki}\ \emph {et~al.}(1996)\citenamefont {Inagaki}, \citenamefont {Koiwai}, \citenamefont {Suzuki}, \citenamefont {Fukushima},\ and\ \citenamefont {Kuroda}}]{Inagaki:1996mc}%
  \BibitemOpen
  \bibfield  {author} {\bibinfo {author} {\bibfnamefont {S.}~\bibnamefont {Inagaki}}, \bibinfo {author} {\bibfnamefont {A.}~\bibnamefont {Koiwai}}, \bibinfo {author} {\bibfnamefont {N.}~\bibnamefont {Suzuki}}, \bibinfo {author} {\bibfnamefont {Y.}~\bibnamefont {Fukushima}},\ and\ \bibinfo {author} {\bibfnamefont {K.}~\bibnamefont {Kuroda}},\ }\bibfield  {title} {\bibinfo {title} {{S}yntheses of {H}ighly {O}rdered {M}esoporous {M}aterials, {FSM}-16, {D}erived from {K}anemite},\ }\href {https://doi.org/10.1246/bcsj.69.1449} {\bibfield  {journal} {\bibinfo  {journal} {Bull. Chem. Soc. Jpn.}\ }\textbf {\bibinfo {volume} {69}},\ \bibinfo {pages} {1449} (\bibinfo {year} {1996})}\BibitemShut {NoStop}%
\bibitem [{\citenamefont {Nichols}\ \emph {et~al.}(2020)\citenamefont {Nichols}, \citenamefont {Prisk}, \citenamefont {Warren}, \citenamefont {Sokol},\ and\ \citenamefont {{Del Maestro}}}]{Nichols:2020of}%
  \BibitemOpen
  \bibfield  {author} {\bibinfo {author} {\bibfnamefont {N.~S.}\ \bibnamefont {Nichols}}, \bibinfo {author} {\bibfnamefont {T.~R.}\ \bibnamefont {Prisk}}, \bibinfo {author} {\bibfnamefont {G.}~\bibnamefont {Warren}}, \bibinfo {author} {\bibfnamefont {P.}~\bibnamefont {Sokol}},\ and\ \bibinfo {author} {\bibfnamefont {A.}~\bibnamefont {{Del Maestro}}},\ }\bibfield  {title} {\bibinfo {title} {{D}imensional reduction of helium-4 inside argon-plated {M{C}M}-41 nanopores},\ }\href {https://doi.org/10.1103/physrevb.102.144505} {\bibfield  {journal} {\bibinfo  {journal} {Phys. Rev. B}\ }\textbf {\bibinfo {volume} {102}},\ \bibinfo {pages} {144505} (\bibinfo {year} {2020})}\BibitemShut {NoStop}%
\bibitem [{\citenamefont {Graham}\ and\ \citenamefont {Taborek}(1989)}]{Graham:1989qh}%
  \BibitemOpen
  \bibfield  {author} {\bibinfo {author} {\bibfnamefont {G.~M.}\ \bibnamefont {Graham}}\ and\ \bibinfo {author} {\bibfnamefont {P.}~\bibnamefont {Taborek}},\ }\bibfield  {title} {\bibinfo {title} {{I}ncomplete wetting of helium {I} on copper},\ }\href {https://doi.org/10.1103/physrevb.40.802} {\bibfield  {journal} {\bibinfo  {journal} {Phys. Rev. B}\ }\textbf {\bibinfo {volume} {40}},\ \bibinfo {pages} {802} (\bibinfo {year} {1989})}\BibitemShut {NoStop}%
\bibitem [{\citenamefont {Rutledge}\ and\ \citenamefont {Taborek}(1992)}]{Rutledge:1992}%
  \BibitemOpen
  \bibfield  {author} {\bibinfo {author} {\bibfnamefont {J.~E.}\ \bibnamefont {Rutledge}}\ and\ \bibinfo {author} {\bibfnamefont {P.}~\bibnamefont {Taborek}},\ }\bibfield  {title} {\bibinfo {title} {Prewetting phase diagram of $^{4}\mathrm{He}$ on cesium},\ }\href {https://doi.org/10.1103/PhysRevLett.69.937} {\bibfield  {journal} {\bibinfo  {journal} {Phys. Rev. Lett.}\ }\textbf {\bibinfo {volume} {69}},\ \bibinfo {pages} {937} (\bibinfo {year} {1992})}\BibitemShut {NoStop}%
\bibitem [{\citenamefont {Cheng}\ \emph {et~al.}(1993{\natexlab{a}})\citenamefont {Cheng}, \citenamefont {Cole}, \citenamefont {Saam},\ and\ \citenamefont {Treiner}}]{Cheng:1993}%
  \BibitemOpen
  \bibfield  {author} {\bibinfo {author} {\bibfnamefont {E.}~\bibnamefont {Cheng}}, \bibinfo {author} {\bibfnamefont {M.~W.}\ \bibnamefont {Cole}}, \bibinfo {author} {\bibfnamefont {W.~F.}\ \bibnamefont {Saam}},\ and\ \bibinfo {author} {\bibfnamefont {J.}~\bibnamefont {Treiner}},\ }\bibfield  {title} {\bibinfo {title} {Wetting transitions of classical liquid films: A nearly universal trend},\ }\href {https://doi.org/10.1103/PhysRevB.48.18214} {\bibfield  {journal} {\bibinfo  {journal} {Phys. Rev. B}\ }\textbf {\bibinfo {volume} {48}},\ \bibinfo {pages} {18214} (\bibinfo {year} {1993}{\natexlab{a}})}\BibitemShut {NoStop}%
\bibitem [{\citenamefont {Cheng}\ \emph {et~al.}(1993{\natexlab{b}})\citenamefont {Cheng}, \citenamefont {Cole}, \citenamefont {Dupont-Roc}, \citenamefont {Saam},\ and\ \citenamefont {Treiner}}]{Cheng:1993ri}%
  \BibitemOpen
  \bibfield  {author} {\bibinfo {author} {\bibfnamefont {E.}~\bibnamefont {Cheng}}, \bibinfo {author} {\bibfnamefont {M.~W.}\ \bibnamefont {Cole}}, \bibinfo {author} {\bibfnamefont {J.}~\bibnamefont {Dupont-Roc}}, \bibinfo {author} {\bibfnamefont {W.~F.}\ \bibnamefont {Saam}},\ and\ \bibinfo {author} {\bibfnamefont {J.}~\bibnamefont {Treiner}},\ }\bibfield  {title} {\bibinfo {title} {{N}ovel wetting behavior in quantum films},\ }\href {https://doi.org/10.1103/revmodphys.65.557} {\bibfield  {journal} {\bibinfo  {journal} {Rev. Mod. Phys.}\ }\textbf {\bibinfo {volume} {65}},\ \bibinfo {pages} {557} (\bibinfo {year} {1993}{\natexlab{b}})}\BibitemShut {NoStop}%
\bibitem [{\citenamefont {Treiner}(1993)}]{Treiner1993}%
  \BibitemOpen
  \bibfield  {author} {\bibinfo {author} {\bibfnamefont {J.}~\bibnamefont {Treiner}},\ }\bibfield  {title} {\bibinfo {title} {Helium mixtures on weak binding substrates},\ }\href {https://doi.org/10.1007/BF00681869} {\bibfield  {journal} {\bibinfo  {journal} {J. Low Temp. Phys.}\ }\textbf {\bibinfo {volume} {92}},\ \bibinfo {pages} {1} (\bibinfo {year} {1993})}\BibitemShut {NoStop}%
\bibitem [{\citenamefont {Vidali}\ \emph {et~al.}(1991)\citenamefont {Vidali}, \citenamefont {Ihm}, \citenamefont {Kim},\ and\ \citenamefont {Cole}}]{Vidali:1991}%
  \BibitemOpen
  \bibfield  {author} {\bibinfo {author} {\bibfnamefont {G.}~\bibnamefont {Vidali}}, \bibinfo {author} {\bibfnamefont {G.}~\bibnamefont {Ihm}}, \bibinfo {author} {\bibfnamefont {H.-Y.}\ \bibnamefont {Kim}},\ and\ \bibinfo {author} {\bibfnamefont {M.~W.}\ \bibnamefont {Cole}},\ }\bibfield  {title} {\bibinfo {title} {Potentials of physical adsorption},\ }\href {https://doi.org/https://doi.org/10.1016/0167-5729(91)90012-M} {\bibfield  {journal} {\bibinfo  {journal} {Surf. Sci. Rep.}\ }\textbf {\bibinfo {volume} {12}},\ \bibinfo {pages} {135} (\bibinfo {year} {1991})}\BibitemShut {NoStop}%
\bibitem [{\citenamefont {Taborek}\ and\ \citenamefont {Rutledge}(1994)}]{Taborek1994}%
  \BibitemOpen
  \bibfield  {author} {\bibinfo {author} {\bibfnamefont {P.}~\bibnamefont {Taborek}}\ and\ \bibinfo {author} {\bibfnamefont {J.}~\bibnamefont {Rutledge}},\ }\bibfield  {title} {\bibinfo {title} {Wetting transitions of helium on weak binding substrates},\ }\href {https://doi.org/10.1016/0921-4526(94)90224-0} {\bibfield  {journal} {\bibinfo  {journal} {Physica B: Cond. Mat.}\ }\textbf {\bibinfo {volume} {197}},\ \bibinfo {pages} {283} (\bibinfo {year} {1994})}\BibitemShut {NoStop}%
\bibitem [{\citenamefont {Van~Cleve}\ \emph {et~al.}(2007)\citenamefont {Van~Cleve}, \citenamefont {Taborek},\ and\ \citenamefont {Rutledge}}]{VanCleve2007}%
  \BibitemOpen
  \bibfield  {author} {\bibinfo {author} {\bibfnamefont {E.}~\bibnamefont {Van~Cleve}}, \bibinfo {author} {\bibfnamefont {P.}~\bibnamefont {Taborek}},\ and\ \bibinfo {author} {\bibfnamefont {J.~E.}\ \bibnamefont {Rutledge}},\ }\bibfield  {title} {\bibinfo {title} {Helium adsorption on lithium substrates},\ }\href {https://doi.org/10.1007/s10909-007-9516-5} {\bibfield  {journal} {\bibinfo  {journal} {J. Low Temp. Phys.}\ }\textbf {\bibinfo {volume} {150}},\ \bibinfo {pages} {1} (\bibinfo {year} {2007})}\BibitemShut {NoStop}%
\bibitem [{\citenamefont {Gatica}\ and\ \citenamefont {Cole}(2009)}]{Gatica:2009yo}%
  \BibitemOpen
  \bibfield  {author} {\bibinfo {author} {\bibfnamefont {S.~M.}\ \bibnamefont {Gatica}}\ and\ \bibinfo {author} {\bibfnamefont {M.~W.}\ \bibnamefont {Cole}},\ }\bibfield  {title} {\bibinfo {title} {{T}o {W}et or {N}ot to {W}et: {T}hat {I}s the {Q}uestion},\ }\href {https://doi.org/10.1007/s10909-009-9885-z} {\bibfield  {journal} {\bibinfo  {journal} {J. Low Temp. Phys.}\ }\textbf {\bibinfo {volume} {157}},\ \bibinfo {pages} {111} (\bibinfo {year} {2009})}\BibitemShut {NoStop}%
\bibitem [{\citenamefont {Nacher}\ and\ \citenamefont {Dupont-Roc}(1991)}]{Nacher:1991}%
  \BibitemOpen
  \bibfield  {author} {\bibinfo {author} {\bibfnamefont {P.~J.}\ \bibnamefont {Nacher}}\ and\ \bibinfo {author} {\bibfnamefont {J.}~\bibnamefont {Dupont-Roc}},\ }\bibfield  {title} {\bibinfo {title} {Experimental evidence for nonwetting with superfluid helium},\ }\href {https://doi.org/10.1103/PhysRevLett.67.2966} {\bibfield  {journal} {\bibinfo  {journal} {Phys. Rev. Lett.}\ }\textbf {\bibinfo {volume} {67}},\ \bibinfo {pages} {2966} (\bibinfo {year} {1991})}\BibitemShut {NoStop}%
\bibitem [{\citenamefont {Ketola}\ \emph {et~al.}(1992)\citenamefont {Ketola}, \citenamefont {Wang},\ and\ \citenamefont {Hallock}}]{Ketola:1992}%
  \BibitemOpen
  \bibfield  {author} {\bibinfo {author} {\bibfnamefont {K.~S.}\ \bibnamefont {Ketola}}, \bibinfo {author} {\bibfnamefont {S.}~\bibnamefont {Wang}},\ and\ \bibinfo {author} {\bibfnamefont {R.~B.}\ \bibnamefont {Hallock}},\ }\bibfield  {title} {\bibinfo {title} {Anomalous wetting of helium on cesium},\ }\href {https://doi.org/10.1103/PhysRevLett.68.201} {\bibfield  {journal} {\bibinfo  {journal} {Phys. Rev. Lett.}\ }\textbf {\bibinfo {volume} {68}},\ \bibinfo {pages} {201} (\bibinfo {year} {1992})}\BibitemShut {NoStop}%
\bibitem [{\citenamefont {Mukherjee}\ \emph {et~al.}(1992)\citenamefont {Mukherjee}, \citenamefont {Druist},\ and\ \citenamefont {Chan}}]{Mukherjee:1992}%
  \BibitemOpen
  \bibfield  {author} {\bibinfo {author} {\bibfnamefont {S.~K.}\ \bibnamefont {Mukherjee}}, \bibinfo {author} {\bibfnamefont {D.~P.}\ \bibnamefont {Druist}},\ and\ \bibinfo {author} {\bibfnamefont {M.~H.~W.}\ \bibnamefont {Chan}},\ }\bibfield  {title} {\bibinfo {title} {Evidence for the absence of 4{H}e adsorption on cesium coated graphite surface},\ }\href {https://doi.org/10.1007/BF00141570} {\bibfield  {journal} {\bibinfo  {journal} {J. Low Temp. Phys.}\ }\textbf {\bibinfo {volume} {87}},\ \bibinfo {pages} {113} (\bibinfo {year} {1992})}\BibitemShut {NoStop}%
\bibitem [{\citenamefont {Ohno}\ \emph {et~al.}(2024)\citenamefont {Ohno}, \citenamefont {Del~Maestro},\ and\ \citenamefont {Lakoba}}]{Lakoba:2024}%
  \BibitemOpen
  \bibfield  {author} {\bibinfo {author} {\bibfnamefont {Y.}~\bibnamefont {Ohno}}, \bibinfo {author} {\bibfnamefont {A.}~\bibnamefont {Del~Maestro}},\ and\ \bibinfo {author} {\bibfnamefont {T.}~\bibnamefont {Lakoba}},\ }\bibfield  {title} {\bibinfo {title} {Efficient simulations of hartree-fock equations by an accelerated gradient descent method},\ }\href {https://doi.org/DOI10.1103/PhysRevE.110.055304} {\bibfield  {journal} {\bibinfo  {journal} {Phys. Rev. E}\ }\textbf {\bibinfo {volume} {110}},\ \bibinfo {pages} {055304} (\bibinfo {year} {2024})}\BibitemShut {NoStop}%
\bibitem [{\citenamefont {Boninsegni}\ \emph {et~al.}(2006{\natexlab{a}})\citenamefont {Boninsegni}, \citenamefont {{Prokof'ev}},\ and\ \citenamefont {Svistunov}}]{Boninsegni:2006ed}%
  \BibitemOpen
  \bibfield  {author} {\bibinfo {author} {\bibfnamefont {M.}~\bibnamefont {Boninsegni}}, \bibinfo {author} {\bibfnamefont {N.}~\bibnamefont {{Prokof'ev}}},\ and\ \bibinfo {author} {\bibfnamefont {B.}~\bibnamefont {Svistunov}},\ }\bibfield  {title} {\bibinfo {title} {{Worm Algorithm for Continuous-Space Path Integral Monte Carlo Simulations}},\ }\href {https://link.aps.org/doi/10.1103/PhysRevLett.96.070601} {\bibfield  {journal} {\bibinfo  {journal} {Phys. Rev. Lett.}\ }\textbf {\bibinfo {volume} {96}},\ \bibinfo {pages} {070601} (\bibinfo {year} {2006}{\natexlab{a}})}\BibitemShut {NoStop}%
\bibitem [{\citenamefont {Boninsegni}\ \emph {et~al.}(2006{\natexlab{b}})\citenamefont {Boninsegni}, \citenamefont {Prokof'ev},\ and\ \citenamefont {Svistunov}}]{Boninsegni:2006gc}%
  \BibitemOpen
  \bibfield  {author} {\bibinfo {author} {\bibfnamefont {M.}~\bibnamefont {Boninsegni}}, \bibinfo {author} {\bibfnamefont {N.~V.}\ \bibnamefont {Prokof'ev}},\ and\ \bibinfo {author} {\bibfnamefont {B.~V.}\ \bibnamefont {Svistunov}},\ }\href {https://doi.org/10.1103/physreve.74.036701} {\bibfield  {journal} {\bibinfo  {journal} {Phys. Rev. E}\ }\textbf {\bibinfo {volume} {74}},\ \bibinfo {pages} {036701} (\bibinfo {year} {2006}{\natexlab{b}})}\BibitemShut {NoStop}%
\bibitem [{\citenamefont {{Del Maestro}}(2024)}]{pimcrepo}%
  \BibitemOpen
  \bibfield  {author} {\bibinfo {author} {\bibfnamefont {A.}~\bibnamefont {{Del Maestro}}},\ }\bibfield  {title} {\bibinfo {title} {{Path Integral Quantum Monte Carlo}},\ }\bibfield  {journal} {\bibinfo  {journal} {{Github Repository}}\ }\href {https://doi.org/10.5281/zenodo.7271914} {10.5281/zenodo.7271914} (\bibinfo {year} {2024})\BibitemShut {NoStop}%
\bibitem [{\citenamefont {Przybytek}\ \emph {et~al.}(2010)\citenamefont {Przybytek}, \citenamefont {Cencek}, \citenamefont {Komasa}, \citenamefont {{\L}ach}, \citenamefont {Jeziorski},\ and\ \citenamefont {Szalewicz}}]{Przybytek:2010js}%
  \BibitemOpen
  \bibfield  {author} {\bibinfo {author} {\bibfnamefont {M.}~\bibnamefont {Przybytek}}, \bibinfo {author} {\bibfnamefont {W.}~\bibnamefont {Cencek}}, \bibinfo {author} {\bibfnamefont {J.}~\bibnamefont {Komasa}}, \bibinfo {author} {\bibfnamefont {G.}~\bibnamefont {{\L}ach}}, \bibinfo {author} {\bibfnamefont {B.}~\bibnamefont {Jeziorski}},\ and\ \bibinfo {author} {\bibfnamefont {K.}~\bibnamefont {Szalewicz}},\ }\bibfield  {title} {\bibinfo {title} {{Relativistic and Quantum Electrodynamics Effects in the Helium Pair Potential}},\ }\href {http://link.aps.org/doi/10.1103/PhysRevLett.104.183003} {\bibfield  {journal} {\bibinfo  {journal} {Phys. Rev. Lett.}\ }\textbf {\bibinfo {volume} {104}},\ \bibinfo {pages} {183003} (\bibinfo {year} {2010})}\BibitemShut {NoStop}%
\bibitem [{\citenamefont {Cencek}\ \emph {et~al.}(2012)\citenamefont {Cencek}, \citenamefont {Przybytek}, \citenamefont {Komasa}, \citenamefont {Mehl}, \citenamefont {Jeziorski},\ and\ \citenamefont {Szalewicz}}]{Cencek:2012iz}%
  \BibitemOpen
  \bibfield  {author} {\bibinfo {author} {\bibfnamefont {W.}~\bibnamefont {Cencek}}, \bibinfo {author} {\bibfnamefont {M.}~\bibnamefont {Przybytek}}, \bibinfo {author} {\bibfnamefont {J.}~\bibnamefont {Komasa}}, \bibinfo {author} {\bibfnamefont {J.~B.}\ \bibnamefont {Mehl}}, \bibinfo {author} {\bibfnamefont {B.}~\bibnamefont {Jeziorski}},\ and\ \bibinfo {author} {\bibfnamefont {K.}~\bibnamefont {Szalewicz}},\ }\bibfield  {title} {\bibinfo {title} {{Effects of adiabatic, relativistic, and quantum electrodynamics interactions on the pair potential and thermophysical properties of helium}},\ }\href {http://scitation.aip.org/content/aip/journal/jcp/136/22/10.1063/1.4712218} {\bibfield  {journal} {\bibinfo  {journal} {J. Chem. Phys.}\ }\textbf {\bibinfo {volume} {136}},\ \bibinfo {pages} {224303} (\bibinfo {year} {2012})}\BibitemShut {NoStop}%
\bibitem [{\citenamefont {Arblaster}(2018)}]{Arblaster2018}%
  \BibitemOpen
  \bibfield  {author} {\bibinfo {author} {\bibfnamefont {J.}~\bibnamefont {Arblaster}},\ }\href {https://books.google.com/books?id=lRBjtgEACAAJ} {\emph {\bibinfo {title} {Selected Values of the Crystallographic Properties of the Elements}}}\ (\bibinfo  {publisher} {ASM International},\ \bibinfo {year} {2018})\BibitemShut {NoStop}%
\bibitem [{\citenamefont {M\"user}\ \emph {et~al.}(1995)\citenamefont {M\"user}, \citenamefont {Nielaba},\ and\ \citenamefont {Binder}}]{Muser1995}%
  \BibitemOpen
  \bibfield  {author} {\bibinfo {author} {\bibfnamefont {M.~H.}\ \bibnamefont {M\"user}}, \bibinfo {author} {\bibfnamefont {P.}~\bibnamefont {Nielaba}},\ and\ \bibinfo {author} {\bibfnamefont {K.}~\bibnamefont {Binder}},\ }\bibfield  {title} {\bibinfo {title} {{Path-integral Monte Carlo study of crystalline Lennard-Jones systems}},\ }\href {https://doi.org/10.1103/PhysRevB.51.2723} {\bibfield  {journal} {\bibinfo  {journal} {Phys. Rev. B}\ }\textbf {\bibinfo {volume} {51}},\ \bibinfo {pages} {2723} (\bibinfo {year} {1995})}\BibitemShut {NoStop}%
\bibitem [{\citenamefont {Batchelder}\ \emph {et~al.}(1967)\citenamefont {Batchelder}, \citenamefont {Losee},\ and\ \citenamefont {Simmons}}]{Batchelder1967}%
  \BibitemOpen
  \bibfield  {author} {\bibinfo {author} {\bibfnamefont {D.~N.}\ \bibnamefont {Batchelder}}, \bibinfo {author} {\bibfnamefont {D.~L.}\ \bibnamefont {Losee}},\ and\ \bibinfo {author} {\bibfnamefont {R.~O.}\ \bibnamefont {Simmons}},\ }\bibfield  {title} {\bibinfo {title} {Measurements of lattice constant, thermal expansion, and isothermal compressibility of neon single crystals},\ }\href {https://doi.org/10.1103/PhysRev.162.767} {\bibfield  {journal} {\bibinfo  {journal} {Phys. Rev.}\ }\textbf {\bibinfo {volume} {162}},\ \bibinfo {pages} {767} (\bibinfo {year} {1967})}\BibitemShut {NoStop}%
\bibitem [{\citenamefont {Boda}\ and\ \citenamefont {Henderson}(2008)}]{Boda:2008}%
  \BibitemOpen
  \bibfield  {author} {\bibinfo {author} {\bibfnamefont {D.}~\bibnamefont {Boda}}\ and\ \bibinfo {author} {\bibfnamefont {D.}~\bibnamefont {Henderson}},\ }\bibfield  {title} {\bibinfo {title} {{The effects of deviations from Lorentz-Berthelot rules on the properties of a simple mixture}},\ }\href {https://doi.org/10.1080/00268970802471137} {\bibfield  {journal} {\bibinfo  {journal} {Mol. Phys.}\ }\textbf {\bibinfo {volume} {106}},\ \bibinfo {pages} {2367} (\bibinfo {year} {2008})}\BibitemShut {NoStop}%
\bibitem [{\citenamefont {Tjatjopoulos}\ \emph {et~al.}(1988)\citenamefont {Tjatjopoulos}, \citenamefont {Feke},\ and\ \citenamefont {Mann}}]{Tjatjopoulos:1988jl}%
  \BibitemOpen
  \bibfield  {author} {\bibinfo {author} {\bibfnamefont {G.~J.}\ \bibnamefont {Tjatjopoulos}}, \bibinfo {author} {\bibfnamefont {D.~L.}\ \bibnamefont {Feke}},\ and\ \bibinfo {author} {\bibfnamefont {J.~A.}\ \bibnamefont {Mann}},\ }\bibfield  {title} {\bibinfo {title} {{M}olecule-micropore interaction potentials},\ }\href {https://doi.org/10.1021/j100324a063} {\bibfield  {journal} {\bibinfo  {journal} {J. Phys. Chem.}\ }\textbf {\bibinfo {volume} {92}},\ \bibinfo {pages} {4006} (\bibinfo {year} {1988})}\BibitemShut {NoStop}%
\bibitem [{\citenamefont {Zhang}\ \emph {et~al.}(2004)\citenamefont {Zhang}, \citenamefont {Wang},\ and\ \citenamefont {Jiang}}]{Zhang:2004xo}%
  \BibitemOpen
  \bibfield  {author} {\bibinfo {author} {\bibfnamefont {X.}~\bibnamefont {Zhang}}, \bibinfo {author} {\bibfnamefont {W.}~\bibnamefont {Wang}},\ and\ \bibinfo {author} {\bibfnamefont {G.}~\bibnamefont {Jiang}},\ }\bibfield  {title} {\bibinfo {title} {{A} potential model for interaction between the {L}ennard–{J}ones cylindrical wall and fluid molecules},\ }\href {https://doi.org/10.1016/j.fluid.2004.01.005} {\bibfield  {journal} {\bibinfo  {journal} {Fluid Phase Equilibr.}\ }\textbf {\bibinfo {volume} {218}},\ \bibinfo {pages} {239} (\bibinfo {year} {2004})}\BibitemShut {NoStop}%
\bibitem [{\citenamefont {Lakoba}\ and\ \citenamefont {Yang}(2007)}]{Lakoba:2007}%
  \BibitemOpen
  \bibfield  {author} {\bibinfo {author} {\bibfnamefont {T.}~\bibnamefont {Lakoba}}\ and\ \bibinfo {author} {\bibfnamefont {J.}~\bibnamefont {Yang}},\ }\bibfield  {title} {\bibinfo {title} {A mode elimination technique to improve convergence of iteration methods for finding solitary waves},\ }\href {https://doi.org/https://doi.org/10.1016/j.jcp.2007.06.010} {\bibfield  {journal} {\bibinfo  {journal} {J. Comput. Phys.}\ }\textbf {\bibinfo {volume} {226}},\ \bibinfo {pages} {1693} (\bibinfo {year} {2007})}\BibitemShut {NoStop}%
\bibitem [{\citenamefont {Ceperley}(1995)}]{Ceperley:1995gr}%
  \BibitemOpen
  \bibfield  {author} {\bibinfo {author} {\bibfnamefont {D.~M.}\ \bibnamefont {Ceperley}},\ }\bibfield  {title} {\bibinfo {title} {Path integrals in the theory of condensed helium},\ }\href {https://doi.org/10.1103/revmodphys.67.279} {\bibfield  {journal} {\bibinfo  {journal} {Rev. Mod. Phys.}\ }\textbf {\bibinfo {volume} {67}},\ \bibinfo {pages} {279 } (\bibinfo {year} {1995})}\BibitemShut {NoStop}%
\bibitem [{\citenamefont {Sarsa}\ \emph {et~al.}(2000)\citenamefont {Sarsa}, \citenamefont {Schmidt},\ and\ \citenamefont {Magro}}]{Sarsa:2000jl}%
  \BibitemOpen
  \bibfield  {author} {\bibinfo {author} {\bibfnamefont {A.}~\bibnamefont {Sarsa}}, \bibinfo {author} {\bibfnamefont {K.~E.}\ \bibnamefont {Schmidt}},\ and\ \bibinfo {author} {\bibfnamefont {W.~R.}\ \bibnamefont {Magro}},\ }\bibfield  {title} {\bibinfo {title} {A path integral ground state method},\ }\href {https://doi.org/10.1063/1.481926} {\bibfield  {journal} {\bibinfo  {journal} {J. Chem. Phys.}\ }\textbf {\bibinfo {volume} {113}},\ \bibinfo {pages} {1366} (\bibinfo {year} {2000})}\BibitemShut {NoStop}%
\bibitem [{\citenamefont {Rosenow}\ and\ \citenamefont {{Del Maestro}}(2024)}]{Rosenow:2024fo}%
  \BibitemOpen
  \bibfield  {author} {\bibinfo {author} {\bibfnamefont {B.}~\bibnamefont {Rosenow}}\ and\ \bibinfo {author} {\bibfnamefont {A.}~\bibnamefont {{Del Maestro}}},\ }\bibfield  {title} {\bibinfo {title} {Friedel oscillations in one-dimensional $^4${He}},\ }\href {https://arxiv.org/abs/2411.13654} {\bibfield  {journal} {\bibinfo  {journal} {arXiv:2411.13654}\ } (\bibinfo {year} {2024})}\BibitemShut {NoStop}%
\bibitem [{\citenamefont {Donnelly}\ and\ \citenamefont {Barenghi}(1998)}]{Donnelly:1998}%
  \BibitemOpen
  \bibfield  {author} {\bibinfo {author} {\bibfnamefont {R.~J.}\ \bibnamefont {Donnelly}}\ and\ \bibinfo {author} {\bibfnamefont {C.~F.}\ \bibnamefont {Barenghi}},\ }\bibfield  {title} {\bibinfo {title} {The observed properties of liquid helium at the saturated vapor pressure},\ }\href {https://doi.org/10.1063/1.556028} {\bibfield  {journal} {\bibinfo  {journal} {J. Phys. Chem. Ref. Data}\ }\textbf {\bibinfo {volume} {27}},\ \bibinfo {pages} {1217} (\bibinfo {year} {1998})}\BibitemShut {NoStop}%
\bibitem [{\citenamefont {{Paul}}(2024)}]{relaxationrepo}%
  \BibitemOpen
  \bibfield  {author} {\bibinfo {author} {\bibfnamefont {S.}~\bibnamefont {{Paul}}},\ }\bibfield  {title} {\bibinfo {title} {\href{https://github.com/paulsphys/relaxation}{Relaxation method}},\ }\bibfield  {journal} {\bibinfo  {journal} {{Github Repository}}\ }\href {https://github.com/paulsphys/relaxation} {} (\bibinfo {year} {2024})\BibitemShut {NoStop}%
\bibitem [{\citenamefont {Paul}\ and\ \citenamefont {{Del Maestro}}(2025{\natexlab{a}})}]{paperrepo}%
  \BibitemOpen
  \bibfield  {author} {\bibinfo {author} {\bibfnamefont {S.}~\bibnamefont {Paul}}\ and\ \bibinfo {author} {\bibfnamefont {A.}~\bibnamefont {{Del Maestro}}},\ }\bibfield  {title} {\bibinfo {title} {{Papers Nanopore Wetting}},\ }\bibfield  {journal} {\bibinfo  {journal} {{Github Repository}}\ }\href {https://doi.org/10.5281/zenodo.17109103} {10.5281/zenodo.17109103} (\bibinfo {year} {2025}{\natexlab{a}})\BibitemShut {NoStop}%
\bibitem [{\citenamefont {Paul}\ and\ \citenamefont {{Del Maestro}}(2025{\natexlab{b}})}]{datarepo}%
  \BibitemOpen
  \bibfield  {author} {\bibinfo {author} {\bibfnamefont {S.}~\bibnamefont {Paul}}\ and\ \bibinfo {author} {\bibfnamefont {A.}~\bibnamefont {{Del Maestro}}},\ }\bibfield  {title} {\bibinfo {title} {{Papers Nanopore Wetting Raw Simulation Data}},\ }\bibfield  {journal} {\bibinfo  {journal} {{Zenodo Data Repository}}\ }\href {https://doi.org/10.5281/zenodo.17108613} {10.5281/zenodo.17108613} (\bibinfo {year} {2025}{\natexlab{b}})\BibitemShut {NoStop}%
\end{thebibliography}%

\end{document}